\documentclass[a4paper,10pt]{article}

\usepackage{epsfig,multicol}
\usepackage{amsmath}
\usepackage{amsfonts}
\usepackage{graphicx}
\usepackage{lscape}
\usepackage{amsthm}
\usepackage{amssymb}
\usepackage{a4}

\theoremstyle{definition}
\newtheorem{defi}{Definition}
\newtheorem{theo}{Theorem}
\newtheorem{coro}{Corollary}

\newcommand{\ee}{\end{eqnarray}}
\newcommand{\be}{\begin{eqnarray}}
\newcommand{\Tr}{\operatorname{Tr}}

\begin{document}

\begin{titlepage}

\begin{flushright}
\small
KUL-TF-06/30\\
\date \\
\normalsize
\end{flushright}
\vspace{1cm}

\begin{centering}

{\huge {\bf Topology of Fibre bundles and Global Aspects of Gauge
Theories}}

\vspace{2cm}

{\large Andr\'es Collinucci\footnote{email: andres.collinucci@fys.kuleuven.be} }\\

\vspace{1cm}

{\small \it Instituut voor Theoretische Fysica, Katholieke
Universiteit Leuven
\\ Celestijnenlaan 200D,
B-3001 Leuven, Belgium \\}

\vspace{1.5cm}

{\large and Alexander Wijns\footnote{Aspirant FWO; e-mail:
awijns@tena4.vub.ac.be}}\\

\vspace{1cm}

{\small \it Theoretische Natuurkunde, Vrije Universitiet Brussel \&
\\International Solvay Institutes, \\ Pleinlaan 2,
B-1050 Brussels, Belgium \\}

\vspace{1.5cm}

\end{centering}

\begin{abstract}
In these lecture notes we will try to give an introduction to the
use of the mathematics of fibre bundles in the understanding of some
global aspects of gauge theories, such as monopoles and instantons.
They are primarily aimed at beginning PhD students. First, we will
briefly review the concept of a fibre bundle and define the notion
of a connection and its curvature on a principal bundle. Then we
will introduce some ideas from (algebraic and differential) topology
such as homotopy, topological degree and characteristic classes.
Finally, we will apply these notions to the bundle setup
corresponding to monopoles and instantons. We will end with some
remarks on index theorems and their applications and some hints
towards a bigger picture.

Lecture presented at the second Modave Summer School on Mathematical
Physics, August 2006.
\end{abstract}

\end{titlepage}

\tableofcontents

\newpage

\section{Fibre Bundles}

A fibre bundle is a manifold\footnote{More generally, one can define
a fibre bundle as being any topological space. We will use the term
fibre bundle in the more restrictive sense of manifolds throughout
these notes. Whenever we will say manifold, we will mean
differentiable manifold.} that looks locally like a product of two
manifolds, but isn't necessarily a product globally. Because of
their importance in modern theoretical physics, many introductory
expositions of fibre bundles for physicists exist. We give a far
from exhaustive list in the references. This section is mainly
inspired by \cite{Nakahara} and \cite{NashandSen}. To get some
intuition for the bundle concept, let us start off with the easiest
possible example.

\subsection{Invitation: the M\"obius strip}
\label{alan:mobius}

\begin{figure}[h]
\begin{center}
\psfig{figure=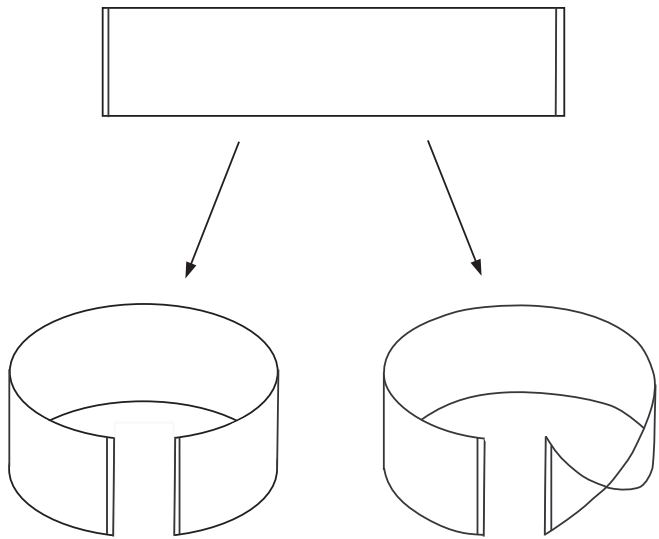,height=3in} \caption{The cylinder and the
M\"obius strip. \label{alan:fig:mobius}}
\end{center}
\end{figure}

Consider a rectangular strip. This can of course be seen as the
product of two line segments. If now one wants to join two opposite
edges of the strip to turn one of the line segments into a circle,
there are two ways to go about this. The first possibility is to
join the two edges in a straightforward way to form a cylinder $C$,
as on the left hand side of figure \ref{alan:fig:mobius}. It should
be intuitively clear that the cylinder is not only locally a
product, but also globally; namely $C = S^1 \times L$, where $L$ is
a line segment. A fibre bundle will be called trivial if it can be
described as a global product. The cylinder is trivial in this
sense, because it is not very difficult to find a global
diffeomorphism from $S^1 \times L$ to $C$.

The second way to join the edges of the strip is of course the more
interesting one. Before gluing the edges together, perform a twist
on one of them to arrive at a M\"obius strip $Mo$, as shown on the
right hand side of figure \ref{alan:fig:mobius}. Locally, along each
open subset $U$ of the $S^1$, the M\"obius strip still looks like a
product, $Mo = U\times L$. Globally, however, there is no
unambiguous and continuous way to write a point $m$ of $Mo$ as a
cartesian pair $(s,\,t) \in S^1 \times L$. The M\"obius strip is
therefore an example of a manifold that is not a global product,
that is, of a non-trivial fibre bundle.

Since $Mo$ is still locally a product, we can try to use this `local
triviality' to our advantage to find a useful way to describe it.
Although we cannot write $Mo$ as $S^1 \times L$, we can still
project down any point $m$ of $Mo$ onto the circle, i.e. there is a
projection $\pi$:
 \be
 \pi: Mo \rightarrow S^1\,,
 \ee
so that, for every $x \in S^1$, its inverse image is isomorphic to
the line segment, $\pi^{-1}(x) \cong L$. This leads to a natural way
to define local coordinates on $Mo$, namely for every open subset
$U$ of $S^1$, we can define a diffeomorphism
 \be
 \phi: U \times L \rightarrow \pi^{-1}(U)\,.
 \ee
This means that to every element $p$ of $\pi^{-1}(U) \subset Mo$, we
can assign local coordinates $\phi^{-1}(p) = (x,\,t)$, where
$x=\pi(p)\in U$ by definition and $t\in L$. Now, how can we quantify
the non-triviality of the M\"obius strip? For this, cover the circle
by two open sets, $U_1$ and $U_2$, which overlap on two disjoint
open intervals, $A$ and $B$. We also have the diffeomorphisms
 \be
 \begin{split}
 &\phi_1: U_1 \times L \rightarrow \pi^{-1}(U_1)\,,\\
 &\phi_2: U_2 \times L \rightarrow \pi^{-1}(U_2)\,.
 \end{split} \label{alan:mobiushom}
 \ee
It is clear that the non-triviality of $Mo$ will reside in the way
in which the different copies of $L$ will be mapped to each other on
$A$ and $B$. To this end, we need an automorphism of $L$ over the
region $A \cup B = U_1 \cap U_2$. This is provided by $\phi_1$ and
$\phi_2$ of eq. (\ref{alan:mobiushom}). For every $x \in U_1 \cap
U_2$, we can define
 \be
 \phi_1^{-1} \circ \phi_2: (A
 \cup B)\times L \rightarrow (A \cup B)\times L\,.
 \ee
This induces a diffeomorphism $g_{12}$ from $L$ to $L$ in such a way
that
 \be
 \phi_1^{-1} \phi_2 (x,\,t) = (x,\, g_{12}(t))\,.
 \ee
Since the only linear diffeomorphisms of $L$ are the identity $e$
and the sign-flip, $g(t) = -t$, we necessarily have that $g_{12} \in
\{ e,\, g \}$. We can always choose $g_{12}=e$ on $A$, so that for
the M\"obius strip $g_{12}$ will have to equal $g$ on $B$. We see
that the non-triviality of the M\"obius strip is encoded in the
non-triviality of these `transition functions'. We can now also
understand the difference with the cylinder in another way. The same
construction for the cylinder would lead to the identity on both $A$
and $B$. So we see that the triviality of the cylinder is reflected
in the triviality of the transition functions and that these
functions encode the non-triviality of the M\"obius strip. Since
$g^2 = e$, in this example the transition functions form the group
$\mathbb{Z}_2$. In general this group will be called the structure
group of the bundle and it will turn out to be an ingredient of
utmost importance in the description of bundles.

\subsection{Definition of a bundle}

Let us now turn to the formal definition of a fibre bundle. Most of
the ingredients should now be intuitively clear from the previous
example.

\begin{defi}
A (differentiable) {\bf fibre bundle} $(E,\,\pi,\,M,\,F,\,G)$ consists of
the following elements:
\renewcommand{\labelenumi}{(\roman{enumi})}
\begin{enumerate}
\item A differentiable manifold $E$ called the {\bf total space}.
\item A differentiable manifold $M$ called the {\bf base space}.
\item A differentiable manifold $F$ called the (typical) {\bf fibre}.
\item A surjection $\pi: E \rightarrow M$ called the {\bf projection}. For
$x \in M$, the inverse image $\pi^{-1}(x)\equiv F_x \cong F$ is
called the fibre at $x$.
\item A (Lie) Group $G$ called the {\bf structure group}, which acts on
the fibre on the left.
\item An open covering $\{U_i\}$ of $M$ and a set of
diffeomorphisms $\phi_i: U_i \times F \rightarrow \pi^{-1}(U_i)$
such that $\pi \phi_i (x,\,t) = x$. The map $\phi_i$ is called a
{\bf local trivialization}. \label{alan:diff}
\item At each point $x\in M$, $\phi_{i,x}(t) \equiv \phi_i(x,\,t)$
is  a diffeomorphism, $\phi_{i,x}: F\rightarrow F_x$. On each
overlap $U_i \cap U_j \neq \emptyset$, we require $g_{ij} = \phi_{i,x}^{-1}
\phi_{j,x}: F\rightarrow F$ to be an element of $G$, i.e. we have a
smooth map $g_{ij}: U_i \cap U_j\rightarrow G$ such that
 \be
 \phi_j(x,\,t) = \phi_i (x,\,g_{ij}(x)t).\nonumber
 \ee
\label{alan:g}
\end{enumerate}
\end{defi}
In the mathematical literature this defines a coordinate bundle. Of
course the properties of a bundle should not depend on the specific
covering of the base manifold or choice of local trivialisations. A
bundle is therefore defined as an equivalence class of coordinate
bundles\footnote{For more details, see for example
\cite{Steenrod}.}. Since in practical applications physicists always
work with an explicit choice of covering and trivialisations, we
will not bother to make this distinction here.

Intuitively, one can view a fibre bundle as a manifold $M$ with a
copy of the fibre $F$ at every point of $M$. The main difference
with a product manifold (trivial bundle) is that the fibres can be
`twisted' so that the global structure becomes more intricate than a
product. This `twisting' is basically encoded in the transition
functions which glue the fibres together in a non-trivial way. For
the M\"obius strip in the previous section, $Mo$ was the total
space, the base space was a circle and the fibre was the line
segment $L$. In that example the structure group turned out to be
the discrete group $\mathbb{Z}_2$. In that respect this was not a
typical example, because in what follows all other examples will
involve continuous structure groups. For convenience we will
sometimes use $E \stackrel{\pi}{\longrightarrow} M$ or simply $E$,
to denote $(E,\,\pi,\,M,\,F,\,G)$.

Let us look at some of the consequences of the above definition.
From (vi) it follows that $\pi^{-1}(U_i)$ is diffeomorphic to a
product, the diffeomorphism given by $\phi_i^{-1}: \pi^{-1}(U_i)
\rightarrow U_i \times F$. It is in this sense that $E$ is locally a
product. From their definition (vii), it is clear that on triple
overlaps, the transition functions obey
 \be
 g_{ij}g_{jk} = g_{ik}\,,\qquad \mbox{on}\quad U_i \cap U_j \cap U_k \neq \emptyset.
 \ee
Taking $i=k$ in the above equation shows that
 \be
 g_{ij}^{-1}=g_{ji} \,,\qquad \mbox{on}\quad U_i \cap U_j \neq \emptyset.
 \ee
These conditions evidently have to be fulfilled to be able to glue
all local pieces of the bundle together in a consistent way. A fibre
bundle is trivial if and only if all transition functions can be
chosen to be identity maps. Since a choice of local trivialization
$\phi_i$ results in a choice of local coordinates, the transition
functions are nothing but a transformation of `coordinates' in going
from one open subset to another. When we will discuss gauge theories
they will represent gauge transformations in going from one patch to
another.

Of course, one should be able to change the choice of local
trivializations (coordinates) within one patch. Say that for an open
covering ${U_i}$ of $M$ we have two sets of trivializations
$\{\phi_i\}$ and $\{\widetilde{\phi}_i\}$ of the same fibre bundle.
Define a map $f_i: F \rightarrow F$ at each point $x \in U_i$
 \be
 f_i(x) = \phi_{i,x}^{-1}\widetilde{\phi}_{i,x}.
 \ee
It is easy to show that the transition functions corresponding to
both trivializations are related by
 \be
 \widetilde{g}_{ij}(x) = f_i(x)^{-1} g_{ij}(x)f_j(x).\label{alan:gauge_f}
 \ee
While the $g_{ij}$ will be gauge transformations for gluing patches
together, the $f_i$ will be gauge transformations within a patch.
From eq. (\ref{alan:gauge_f}) it's clear that in general the
transition functions of a trivial bundle will have the factorized
form
 \be
 g_{ij}(x) = f_i(x)^{-1}f_j(x).
 \ee

\subsection{More examples: vector and principal bundles}
\label{alan:bundle_ex}

The prototype of a fibre bundle is the tangent bundle of a
differentiable manifold. As described in subsection
\ref{alan:manifolds}, the collection of all tangent vectors to a
manifold $M$ at a point $x$ is a vector space called the tangent
space $T_xM$. The collection $\{T_xM \vert x \in M\}$ of all tangent
spaces of $M$ is called the tangent bundle $TM$. Its base manifold
is $M$ and fibre $\mathbb{R}^m$, where $m$ is the dimension of $M$.
Its structure group is a subgroup of $GL(m,\mathbb{R})$. Let us look
at some examples of tangent bundles.

\newtheorem*{tr_n}{$T\mathbb{R}^n$}
\begin{tr_n}
If $M$ is $\mathbb{R}^n$, the tangent space to every point is
isomorphic to $M$ itself. Its tangent bundle $T\mathbb{R}^n$ is
clearly trivial and equal to $\mathbb{R}^n \times \mathbb{R}^n \cong
\mathbb{R}^{2n}$. It can be proven that every bundle over a manifold
that is contractible to a point is trivial.
\end{tr_n}

\newtheorem*{ts1}{$TS^1$}
\begin{ts1}
The circle is not contractible, yet its tangent bundle $TS^1$ is
trivial. The reason is that since one can globally define a (unit)
vector along the circle in an unambiguous and smooth way, it is easy
to find a diffeomorphism from $TS^1$ to $S^1 \times \mathbb{R}$.
\end{ts1}

\newtheorem*{ts2}{$TS^2$}
\begin{ts2}
The tangent bundle of the 2-sphere $TS^2$ is our second example of a
non-trivial bundle. There is no global diffeomorphism between $TS^2$
and $S^2 \times \mathbb{R}^2$, since to establish this one would
have to be able to define two linearly independent vectors at every
point of the sphere in a smooth fashion. (This is needed to be able
to define coordinates on the tangent plane in a smooth way along the
sphere.) In fact it's even worse for the 2-sphere, since in this
case one cannot even find a single global non-vanishing vector
field. The fact that this isn't possible has become known as the
expression: ``You cannot comb hair on a sphere."
\end{ts2}

\newtheorem*{fs2}{$FS^2$}
\begin{fs2}
A set of pointwise linearly independent vectors over an open set of
the base manifold of a tangent bundle is called a frame. So in the
example above, the non-triviality of $TS^2$ was a consequence of not
being able to find a frame over the entire sphere in a consistent
way. At each point one can of course construct many different sets
of linearly independent vectors. These are all related to each other
by a transformation in the structure group, $GL(2,\mathbb{R})$ in
this case. Since the action of $GL(2,\mathbb{R})$ on the set of
frames is free (no fixed points for $g\neq e$) and transitive (every
frame can be obtained from a fixed reference frame by a group
element), the set of all possible frames over an open set $U$ of
$S^2$ is diffeomorphic to $U \times GL(2,\mathbb{R})$. Globally this
becomes a bundle over $S^2$ with fibre $GL(2,\mathbb{R})$ and is
called the frame bundle $FS^2$ of $S^2$. Note that for this bundle
the fibre equals the structure group!
\end{fs2}

The first three examples above are examples of vector bundles, so
let us define these properly.
\begin{defi}
A {\bf vector bundle} $E \stackrel{\pi}{\longrightarrow} M$ is a
fibre bundle whose fibre is a vector space. If $F = \mathbb{R}^n$ it
is common to call $n$ the fibre dimension and denote it by
$\mbox{dim} E$ (although the total dimension of the bundle is
$\mbox{dim} M+n$). The transition functions belong to
$GL(n,\mathbb{R})$.
\end{defi}

Clearly a tangent bundle is always a vector bundle. Once one defines
a frame $\{e_a\}$, $a \in \{1,...,n\}$, over a patch $U \subset M$,
one can expand any vector field $V: U \rightarrow \mathbb{R}^n$ over
$U$ in terms of this frame, $V = V^a e_a$. A possible local
trivialization would then become
 \be
 \phi^{-1}(p) = (\pi(p),\{V^a\})\,, \qquad p \in E
 \ee
Consider two coordinate frames, associated to a set of coordinates
$\{x^a\}$ on $U_x$ and $\{y^a\}$ on $U_y$ respectively. A vector
field $V$ on the overlap $U_x \cap U_y$ can be expanded using either
frame (this notation is discussed in more detail in subsection
\ref{alan:manifolds}),
 \be
 V = V^a \frac{\partial}{\partial x^a} = \widetilde{V}^a \frac{\partial}{\partial y^a}
 \ee
The resulting trivializations are related as follows (in a hopefully
obvious notation),
 \be
 \phi_y^{-1} \phi_x (\pi(p),\,\{V^a\}) = (\pi(p),\{\widetilde{V}^a\}),
 \ee
where
 \be
 V^a = \frac{\partial x^a}{\partial y^b} \widetilde{V}^b.
 \ee
This relation is of course well known from basic tensor calculus. In
the language we developed in the previous section this would be
written as
 \be
 \left(g_{yx}\right)^a_{~b} = \frac{\partial x^a}{\partial y^b}.
 \ee

In the frame bundle example above, the fibre was not a vector space,
but a Lie group. More importantly, the fibre equalled the structure
group. This is an example of a principal bundle, the most important
kind of bundle for understanding the topology of gauge theories. We
will need them a lot in this set of lectures, so let us look at them
in a little more detail.

\begin{defi}
A {\bf principal bundle} has a fibre which is identical to the
structure group $G$. It is usually denoted by $P(M,\,G)$ and called
a $G$-bundle over $M$.
\end{defi}

Most of the time, $G$ will be a Lie group. The only example we will
encounter where this is not the case, is the M\"obius strip. The
action of the structure group on the fibre now simply becomes left
multiplication within $G$. In addition, we can also define right
multiplication (on an element $p$ of $P$) as follows. Let $\phi_i:
U_i \times G \rightarrow \pi^{-1}(U_i)$ be a local trivialization,
 \be
 \phi_i^{-1}(p) = (x,\,g_i),\qquad x = \pi(p).
 \ee
Right multiplication by an element $a$ of $G$ is defined by
 \be
 p\,a = \phi_i(x,\,g_ia).
 \ee
Since left and right multiplications commute (associativity of the
group), this action is independent of the choice of local
coordinates. Let $x \in U_i \cap U_j$, then
 \be
 p\,a = \phi_j(x,\,g_ja) = \phi_j(x,\,g_{ji}(x)g_i a) =
 \phi_i(x,\,g_ia).
 \ee
We can thus just as well write the action as $P\times G \rightarrow
P: (p,\,a) \mapsto p\,a$, without reference to local choices. One
can show that this action is transitive and free on $\pi^{-1}(x)$
for each $x\in M$.

\subsection{Triviality of a bundle}
\label{alan:triviality}

Later on, we will discuss a number of ways in which one can quantify
the non-triviality of a given bundle. Of course before we try to
quantify how much it deviates from triviality, it is interesting to
know wether it is non-trivial at all. We will now discuss some
equivalent ways to see wether a bundle is a global product or not.
Before we do that, we define one more important notion

\begin{defi}
A {\bf section} $s$ is a smooth map $s: M \rightarrow E$ such that $\pi
s(x) = x$ for all $x \in M$. This is sometimes also referred to as a
global section. If a section can only be defined on an open set $U$
of $M$, it is called a {\bf local section} and one only has the smooth map
$s: U \rightarrow E$. The set of all sections of $E$ is called
$\Gamma(M,\,E)$, while the set of all local sections over $U$ will
be called $\Gamma(U,\,E)$.
\end{defi}

The best known example of this is a vector field over a manifold
$M$, which is a section of the tangent bundle $TM$. Clearly, it is
not a great challenge to construct a local section over some open
subset $U \subset M$. Being able to construct a (global) section
over $M$ will place much stronger requirements on the topology of
the bundle and it will have a lot to say about the non-triviality of
a bundle. This is reflected in the following theorem:

\begin{theo}
A vector bundle of rank $n$ is trivial if and only if it admits $n$
point-wise linearly independent sections, i.e. a global frame.
\label{alan:vectortrivial}
\end{theo}

This is precisely why the tangent bundle over the 2-sphere is
non-trivial. On the other hand, for a vector bundle there always
exists at least one global section, namely the so called zero
section, i.e. the section which maps every point $x \in M$ to
$(x,\,O)$, where $O$ is the origin of the vector space. This is
always possible irrespective of the non-triviality of the bundle,
since we do not need a frame to characterize the origin uniquely.
This is not so for a principal bundle and the group structure of the
fibre allows for the following very powerful theorem:

\begin{theo}
A principal bundle is trivial if and only if it allows a global
section. \label{alan:principaltrivial}
\end{theo}

\begin{figure}[h]
\begin{center}
\psfig{figure=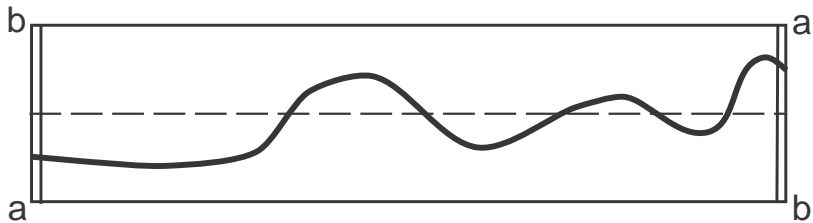,height=1.5in} \caption{A section of the
bundle corresponding to the M\"obius strip. The points marked with
the same letter ($a$ or $b$) should be identified.
\label{alan:fig:section}}
\end{center}
\end{figure}

To illustrate the above theorems, we return to the discussion of the
M\"obius strip $Mo$. In the first section we saw that the
non-triviality of the M\"obius strip had to do with the fact that
one could not find a global trivialization. We now understand that
this is the case because one cannot define a linearly independent
(which in one dimension means everywhere nonzero) section on
$Mo$\footnote{The reader might be bothered by the fact that the
M\"obius strip is not a vector bundle. One can however replace the
line segment $L$ by the real line $\mathbb{R}$ and all the arguments
used in the text will still apply. In this case one would speak of a
line bundle over $S^1$}. It is not very difficult to see (figure
\ref{alan:fig:section}) that every smooth section would have to take
the value zero over at least one point of the circle. According to
Theorem \ref{alan:vectortrivial} this is equivalent to the bundle
being non-trivial.

\begin{figure}[h]
\begin{center}
\psfig{figure=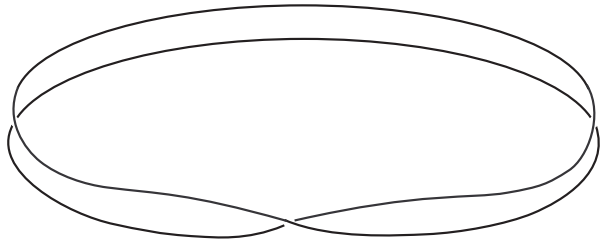,height=1.5in} \caption{The principal
bundle $P(S^1,\mathbb{Z}_2)$ associated to the M\"obius strip is a
double cover of the circle. \label{alan:fig:doublecover}}
\end{center}
\end{figure}

To illustrate the second theorem, we would like to associate a
principal bundle $P(S^1,\mathbb{Z}_2)$ to $Mo$. To accomplish this,
replace the fibre $L$ by the structure group $\mathbb{Z}_2$. Take
the same open covering of the circle as in subsection
\ref{alan:mobius} and use the same transition functions on the
overlap $A \cup B$, where now instead of acting on the line segment
$L$, they act by left multiplication within $\mathbb{Z}_2$. What one
gets is a double cover of the circle as depicted in figure
\ref{alan:fig:doublecover}. To get a global section one has to go
around the circle once and it is clear that this will always have a
discontinuity somewhere. Thus we cannot find a section of this
principal bundle, so that according to Theorem
\ref{alan:principaltrivial} it has to be non-trivial.

The above construction is more general: starting from a fibre bundle
(not necessarily a vector bundle), one can always construct the
associated principal bundle by replacing the fibre with the
structure group and keeping the transition functions. Note that the
frame bundle $FS^2$ over $S^2$ from subsection \ref{alan:bundle_ex}
was nothing but the principal bundle associated to the tangent
bundle $TS^2$. More generally two bundles with same base space and
structure group are called associated if their respective associated
principal bundles are equivalent. In gauge theories it is important
to be able to associate a principal bundle to a vector bundle and
vice versa.

\begin{defi}
Start from a principal bundle $P(M,\,G)$ and an $n$-dimensional
faithful representation $\rho: G \rightarrow GL(n,\,\mathbb{R})$
which acts on $V = \mathbb{R}^n$ from the left. Consider the product
$P\times V$. Define the equivalence relation $(p,\,v)\sim
(p\,g^{-1}, \,\rho(g)v)$, where $p\in P$, $v\in V$ and $g\in G$. The
vector bundle $E_\rho$ associated to $P$ via the representation
$\rho$ is defined as
 \be
 E_\rho = P \times_\rho V \equiv P\times V/ \sim
 \ee
\end{defi}

This is basically a complicated way of saying that one changes the
fibre from $G$ to $V$ and use as transition function $\rho(g_{ij})$
instead of $g_{ij}$. Since every element of $P$ over a certain point
$x\in M$ can be obtained from $(x,\,e)$ (where $e$ is the identity
of $G$) by an element of $G$, the equivalence relation
$(p\,g,\,v)\sim (p,\,\rho(g)v)$ effectively replaces the fibre over
$x$ with $V$, thus replacing a principal bundle by a vector bundle.
If we define the new projection by
 \be
 \pi_E[(p,\,v)] = \pi(p),
 \ee
this is well defined under the equivalence relation since $\pi(p\,
g)=\pi(p)$. If $\phi_P(\pi(p),\,g) = p$, $p\in P$ is a local
trivialization on $U\subset M$, we define for $E_\rho$
 \be
 \phi_E^{-1}: E_\rho \rightarrow U\times V: [(p,\,v)] \mapsto
 (\pi(p),\,\rho(g)v).
 \ee
This definition is independent of the representative of the
equivalence class. To see this, take two different representatives:
 \be
 [(p,\,v)] = [(p\,h^{-1},\,\rho(h)v)].
 \ee
To find their trivialization corresponding to a trivialization over
$U$ in the associated principal bundle $P$, we note that
 \be
 p = \phi_P(x,g), && x = \pi(p);\\
 p\,h^{-1} = \phi_P(x,g\,h^{-1}), && x = \pi(p\,h^{-1})=\pi(p).
 \ee
From this, we find
 \be
 [(p\,h^{-1},\,\rho(h)v)] = \phi_E (x,\,\rho(g\,h^{-1})\rho(h)v) = \phi_E
 (x,\,\rho(g)v).
 \ee
To see what the transition functions are, consider a point
$[(p,\,v)]\in E_\rho$. If we choose a trivialization of $U_i$ such
that $p = \phi_P^i(x,\,e)$, then on $U_j$ there is a trivialization
such that on $U_i \cap U_j \neq \emptyset$, we have $p =
\phi_P^i(x,\, g_{ji})$. For the corresponding trivializations of
$E_\rho$, this means
 \be
 [(p,\,v)] = \phi^i_E(x,v) = \phi^j_E(x,\rho(g_{ji})v) = \phi^j_E(x,\rho_{ji}v).
 \ee
This shows that the new transition functions $\rho_{ij}$ are just
$\rho(g_{ij})$.

Now that we have defined all necessary ingredients, we can formulate
the main theorem of this subsection\footnote{There is actually a
more general theorem which states that a bundle is trivial iff all
its associated bundles are trivial.}

\begin{theo}
A vector bundle is trivial if and only if its associated principal
bundle is trivial.
\end{theo}

This means that for establishing the (non-)triviality of a vector
bundle, we only need to study its associated principal bundle. More
concretely:

\begin{coro}
A vector bundle is trivial if an only if its associated principal
bundle admits a section.
\end{coro}

Sometimes the converse is more useful:

\begin{coro}
A principal bundle is trivial if an only if its associated vector
bundle of rank $n$ admits $n$ point-wise linearly independent
sections.
\end{coro}

This is exactly why both the bundle $Mo$ and its associated
principal bundle were non-trivial. We were basically looking at the
same issue from two different points of view.

\section{Connections on fibre bundles}

Now that we have gained some feeling for the concept of a bundle, we
want to define some extra structure on it. We are all familiar with
the notion of parallel transport of a tangent vector on a manifold
in General Relativity. Given a curve in space-time, there are many
possible choices to transport a given vector along this curve, which
are all equally valid as a choice for what `parallel' might
mean\footnote{The Levi-Civita connection corresponds to a choice of
parallel transport such that if a curve is the shortest path between
two points, a tangent vector to this curve stays tangent to the
curve under parallel transport.}. Translated into the language of
fibre bundles, we want to, given a curve $\gamma$ in the base
(space-time) $M$, define a corresponding section of the tangent
bundle $s^\gamma\in TM$, in such a way that $\pi(s^\gamma)=\gamma$
(otherwise it would not be a section).

The question we want to ask ourselves now is basically a
generalization of this. Given a certain motion in the base manifold,
how can we define a corresponding motion in the fibre? Since, given
a bundle, there is no a priori notion of what `parallel' should
mean, we need some additional structure to give meaning to the
notion of `parallel motion'. This will be provided by the choice of
a connection on the bundle. This section follows parts of
\cite{Nakahara} and \cite{Bertlmann} closely, although some specific
points are more indebted to \cite{Frankel}. We will start by
defining parallel transport on a principal bundle and later sketch
how this connection can be used to provide a connection on an
associated vector bundle. But first of all, let us pause and recall
some facts about Lie groups.

\subsection{Lie groups and algebras}
\label{alan:lie}

From now on the structure group $G$ will be a Lie group, i.e. a
differential manifold with a group structure, where the group
operations (multiplication, inverse) are differentiable. Given an
element $g\in G$, we can define the left- and right-translation of
an element $h\in G$ by $g$,
 \be
 L_g (h) &=& g h; \\
 R_g (h) &=& h g.
 \ee
These induce differential maps in the tangent space (see subsection
\ref{alan:pushandpull} for a review of differential maps)
 \be
 L_{g*}:&& T_hG\rightarrow T_{gh}G;\\
 R_{g*}:&& T_hG\rightarrow T_{hg}G.
 \ee
We say that a vector field $X$ is left-invariant if it satisfies
 \be
 L_{g*}\left(X\vert_h \right) = X\vert_{gh}.\label{alan:leftinv}
 \ee
One can show that if $X$ and $Y$ are left-invariant vector fields,
their Lie bracket $[X,\,Y]$ is also left-invariant. The algebra
formed in this way by the left-invariant vector fields of $G$, is
called the Lie algebra $\mathfrak{g}$. Because of
(\ref{alan:leftinv}), a vector $A$ in $T_eG$ ($e$ is the unit
element of $G$) uniquely defines a left-invariant vector field $X_A$
over $G$ (that is, a section of $TG$). This establishes an
isomorphism between $T_eG$ and $\mathfrak{g}$. From now on we will
not make the distinction between the two anymore and say that the
Lie algebra $\mathfrak{g}$ is the tangent space to the identity in
$G$. The generators $T_a$, $a\in\{1,...,\,r\}$ ($r =
\mbox{dim}\mathfrak{g}=\mbox{dim}G$) of $\mathfrak{g}$ satisfy the
well known relations
 \be
 [T_a,T_b]= f_{ab}^{~~c}\,T_c\,,
 \ee
where $f_{ab}^{~~c}$ are the structure constants of $\mathfrak{g}$
(and can be shown to really be constant).

By Lie group we will always mean a matrix group (subgroup of $GL(n)$
with matrix multiplication as group operation), although most of
what we will discuss can be proven in a more general context. From
our experience with matrix groups we know that exponentiation maps
elements of the Lie algebra to elements of the Lie group.
Consequently, $A\in\mathfrak{g}$ generates a curve (one-parameter
subgroup) through $g$ in $G$ by
 \be
 \sigma_t(g) = g \exp(t A) = R_{\exp(t A)}(g).
 \ee
The corresponding flow equation is (in matrix notation)
 \be
 \left.\frac{d \sigma_t(g)}{dt}\right\vert_{t=0} = g A = L_{g*}A =
 X_ A\vert_g\,, \label{alan:flow_eq}
 \ee
where we used that in matrix notation $L_{g*}A = g A$ (exercise) and
that $A$ generates a left-invariant vector field $X_A$. This shows
that the tangent vector to the curve through $g$ is nothing but the
left-translation of $A$ by $L_{g*}$ (or the left-invariant vector
field generated by $A$ evaluated at $g$, $X_A\vert_g$). More
formally one would define the tangent vector to the one-parameter
flow by making use of a function $f:G\rightarrow \mathbb{R}$,
 \be
 X_A (f(g)) = \left.\frac{d}{dt}f(\sigma_t(g))\right\vert_{t=0}.
 \label{alan:formalflow}
 \ee

Given a basis $\{T_a\}$ of $\mathfrak{g}$, one defines a
corresponding basis $\{X_a\}$ of $T_gG$ by left translation to $g\in
G$, $X_a = L_{g*}T_a$. This means that for an element
$A\in\mathfrak{g}$, if $A = A^a T_a$, the left-invariant vector
field corresponding to $A$ can be expanded as $X_A = A^a X_a$. One
can also define a basis for left-invariant 1-forms $\{\eta^a\}$ dual
to to $\{X_a\}$ by, $\eta^a (X_b) = \delta^a_b$. The Maurer-Cartan
form $\Theta$ is then defined by
 \be
 \Theta = T_a \otimes \eta^a. \label{alan:maurer-cartan_formal}
 \ee
This is a $\mathfrak{g}$-valued 1-form, which takes a left-invariant
vector field $X\vert_g$ at $g$ and pulls it back to the identity,
giving back the Lie algebra element $X\vert_e$. To see this,
evaluate the Maurer-Cartan form on $X_A$ defined above,
 \be
 \Theta(X_A)\vert_g = T_a \otimes \eta^a(A^b X_b) = A^a T_a = A.
 \ee
In this way, the Maurer-Cartan form establishes an explicit
isomorphism between $\mathfrak{g}$ and $T_eG$. More generally, it
takes any vector at $g$ and returns a Lie algebra element, thus
establishing the decomposition of the vector in terms of a basis of
left-invariant vector fields at $g$.

To make contact with widely used notation in physics, we will now be
a bit less formal. Choosing coordinates $\{g^i\}$ on a patch of $G$,
a coordinate basis at $g$ can be written as $\{\partial/\partial
g^i\}$. A coordinate basis at the identity $e$ would in this
notation be written as
 \be
 \left. \frac{\partial}{\partial g^i}\right\vert_{e} = L_{g^{-1}*}\frac{\partial}{\partial
 g^i}.
 \ee
This would mean that (\ref{alan:maurer-cartan_formal}) can be
rewritten as
 \be
 \Theta = L_{g^{-1}*}\frac{\partial}{\partial g^i} \otimes dg^i =
 g^{-1} \frac{\partial}{\partial g^i} \otimes dg^i,
 \ee
where the last equality is for a matrix group. Physicists write this
as
 \be
 \Theta = g^{-1} d g , \label{alan:maurer-cartan_practical}
 \ee
where $dg$ should be interpreted as the identity operator at $g$,
 \be
 d g = \frac{\partial}{\partial g^i} \otimes dg^i.
 \ee
Since for a matrix group $X_A\vert_g = gA$, we get indeed that
 \be
 \Theta(X_A) = g^{-1} d g(X_A) = g^{-1} g A = A.
 \ee
The reason why this notation makes sense is because if $A$ is
tangent to a flow $\sigma_t(e)$, $X_A\vert_g = gA$ is tangent to the
flow $\sigma_t(g)$. Concretely, we have
 \be
 A = \left. \frac{d \sigma_t(e)}{dt}\right\vert_{t=0},
 \ee
so that,
 \be
 X_A\vert_g = g A = \left.\frac{d \sigma_t(g)}{dt}\right\vert_{t=0} \equiv
 X_A(g) = d g(X_A).
 \ee
We see that if we interpret $X_A(g)$ as $X_A\vert_g$ (for a matrix
group in matrix notation), we can almost act as if $dg(X_A)$ has the
usual meaning. In a lot of practical calculations
(\ref{alan:maurer-cartan_practical}) is used in an even more direct
sense: if $g = \exp tA$ for some $A\in G$ and some $t$,
 \be
 \Theta\vert_g = h^{-1}dh\vert_g \equiv \left. e^{-tA} \frac{d}{dt'}e^{t'A}\right\vert_{t'=t} = A,
 \ee
where $\Theta\vert_g$ should now be thought of as simply the Lie
algebra element associated to $g$. What we are doing here is exactly
the same as above, because
 \be
 \left. \frac{d}{dt'}e^{t'A}\right\vert_{t'=t} = e^{tA}A = g A = g
 \left. \frac{d}{dt'}e^{t'A}\right\vert_{t'=0},
 \ee
so that
 \be
 \Theta\vert_g = \left. e^{-tA}
 \frac{d}{dt'}e^{t'A}\right\vert_{t'=t} = g^{-1} \left. \frac{d}{dt'}g
 e^{t'A}\right\vert_{t'=0} = \Theta(X_A),
 \ee
where $X_A$ is the left invariant vector field associated to $A$.

Finally the adjoint map
 \be
 ad_g: G \rightarrow G: h \mapsto ghg^{-1},
 \ee
induces a map in the tangent space
 \be
 ad_{g*}: T_hG \rightarrow T_{ghg^{-1}}G.
 \ee
When we restrict this map to $h=e$, we get the adjoint
representation of $T_eG \cong \mathfrak{g}$,
 \be
 Ad_g: \mathfrak{g}\rightarrow\mathfrak{g}: V \mapsto gVg^{-1}.
 \ee

\subsection{Parallel transport in a principal bundle}
\label{alan:parallel}

\begin{figure}[h]
\begin{center}
\psfig{figure=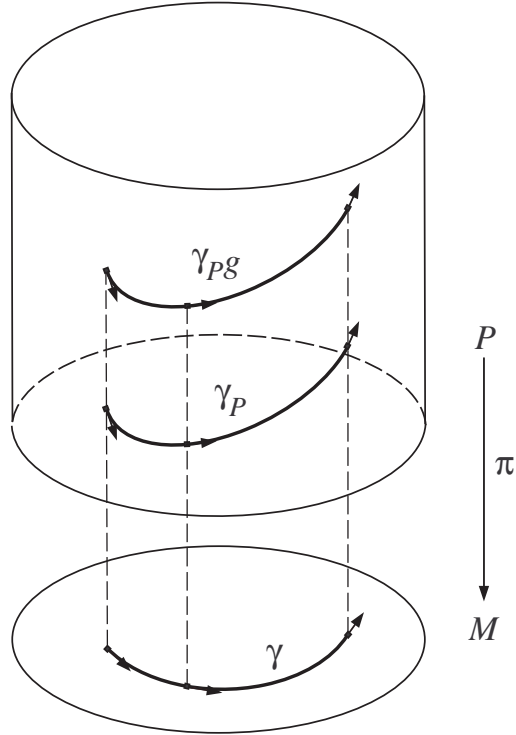,height=4in} \caption{Illustration
of horizontal lift. \label{alan:fig:horizontal_lift}}
\end{center}
\end{figure}

Consider a principal bundle $P(M,\,G)$. Given a curve $\gamma$ on
the base manifold $M$, to define parallel transport, we want to
define a corresponding choice of curve $\gamma_P$ in the total space
$P$. There are of course many possible choices. To characterize our
choice, we will look at vectors tangent to this curve. At every
point along $\gamma$, we can define a lift of the tangent vector to
$\gamma$ to an element in $TP$, the tangent vectors to $P$. This
will define an integral curve $\gamma_P$ in $P$. See figure
\ref{alan:fig:horizontal_lift} for an illustration

The question is then how to lift a vector in $T_\gamma M$ to $TP$.
At every point $p\in P$, we can decompose $T_pP$ into a subspace of
vectors tangent to the fibre $G$, called the vertical subspace
$V_pP$ and a complement $H_pP$, called the horizontal subspace, such
that $T_pP = V_pP \oplus H_pP$. Since $V_pP$ corresponds to motion
along the fibres and is essentially fixed, a choice of $H_pP$ is the
crucial ingredient in the definition of parallel transport. We will
require the vectors tangent to $\gamma_P$ to lie in $H_pP$.

A choice of connection now essentially boils down to a choice of
horizontal subspace. Let us be a bit more precise.

\begin{defi}
A {\bf connection} on $P$ is a smooth and unique separation of the tangent
space $T_pP$ at each $p$ into a vertical subspace $V_pP$ and a
horizontal subspace $H_pP$ such that
\renewcommand{\labelenumi}{(\roman{enumi})}
\begin{enumerate}
\item $T_pP = V_pP \oplus H_pP$; \label{alan:directsum}
\item $H_{pg}P = R_{g*}H_pP$ for every $g\in G$. \label{alan:equivariant}
\end{enumerate}
\end{defi}

Condition (i) just means that every $X\in T_pP$ can be written in a
unique way as a sum $X=X^V + X^H$, where $X^V\in V_pP$ and $X^H\in
H_pP$. The equivariance condition (ii) means that the choice of
horizontal subspace at $p$ determines all the horizontal subspaces
at points $pg$. This roughly means that all points above the same
point $x=\pi(p)$ in the base space will be parallel transported in
the same way (recall that $\pi(p)=\pi(pg)$).

Parallel transport can now immediately be defined by what is called
a horizontal lift.

\begin{defi}
Let $\gamma : [0,\,1]\rightarrow M$ be a curve in the base manifold
(a base curve). A curve $\gamma_P: [0,\,1]\rightarrow P$ is called
the {\bf horizontal lift} of $\gamma$ if
\renewcommand{\labelenumi}{(\roman{enumi})}
\begin{enumerate}
\item $\pi(\gamma_P)=\gamma$;
\item All tangent vectors $X_P$ to $\gamma_P$ are horizontal: $X_P\in
H_{\gamma_P}P$.
\end{enumerate}
\end{defi}

\begin{theo}
Let $\gamma : [0,\,1]\rightarrow M$ be a base curve and let $p\in
\pi^{-1}(\gamma(0))$. Given a connection, there exists a unique
horizontal lift $\gamma_P$ such that $\gamma_P(0)=p$.
\end{theo}

This means that we can (given a connection) uniquely define the
parallel transport of a point $p$ in $P$ along a curve $\gamma$ in
$M$ by moving it along the unique horizontal lift of $\gamma$
through $p$.

A loop in $M$ is defined as a curve $\gamma$ with $\gamma(0)=
\gamma(1)$. It is interesting to see what happens to a horizontal
lift of this loop. In other words, what happens if we parallel
transport an element of $P$ along a closed loop? Starting from a
point $p$ and moving it along a horizontal lift of a loop, there is
no guarantee that we will end up at the same point. In general we
will obtain a different point $p_\gamma$, which depends on the loop
$\gamma$. Since
 \be
 \pi(\gamma_P(0)) = \gamma(0) = \gamma(1) = \pi(\gamma_P(1)),
 \ee
we know that both points will belong to the same fibre, $\pi(p)=
\pi(p_\gamma)$. This means that $p_\gamma = p g$ for some $g\in G$.
If we vary the loop $\gamma$, but keep the base point $p$ fixed, we
generate a group called the holonomy group $Hol_p(P)$ of $P$ at $p$,
which by definition is a subgroup of $G$. This group of course also
depends on the connection, so $Hol_p(P)$ is a characteristic not
only of $P$, but also of the connection. If $M$ is connected the
holonomy group at all points of $M$ are isomorphic and we can speak
of the holonomy group of $P$, $Hol(P)$.

Given a notion of parallel transport in a principal bundle $P$, one
can easily define parallel transport in an associated bundle
$E_\rho$ by

\begin{defi}
If $\gamma_P(t)$ is a horizontal lift of $\gamma(t)\in M$ in $P$,
then $\gamma_E(t)$ is defined to be the horizontal lift of
$\gamma(t)$ in $E_\rho$ if
 \be
 \gamma_E(t) = [(\gamma_P(t),v)],
 \ee
where $v$ is a constant element of $V$.
\end{defi}

This is independent of the lift chosen in $P$, since (because of
equivariance) another lift would be related to $\gamma_P$ by
$\gamma_P'(t) = \gamma_P(t) a$, with a constant element $a\in G$, so
that
 \be
 \gamma_E'(t) = [(\gamma_P'(t),v)] = [(\gamma_P(t),\rho(a)^{-1}v)]
 \ee
where $\rho(a)^{-1}v$ is still a constant element. So this would
still be a horizontal lift, albeit through a different element of
$V$. Choosing a trivialization for $\gamma_P$, $\gamma_P(t) =
\phi_P(\gamma(t),g(t))$, leads to the corresponding trivialization
for $\gamma_E$
 \be
 \gamma_E(t) = \phi_E(\gamma(t),\rho(g(t))v).
 \ee
We see that if parallel transport in $P$ is described by $g(t)$,
then parallel transport in $E_\rho$ is defined by $\rho(g(t))$.

\subsection{Connection one-form on a principal bundle}

Up to this point, the reader might be confused as to what this all
has to do with gauge theories and the usual definition of a
connection in physics. To establish the link with physics, we now
introduce the connection one-form and clarify its relation to the
gauge potential in Yang-Mills theories.

To define the connection one-form properly, we need a more specific
construction of the vertical subspace $V_pP$. Let $A\in
\mathfrak{g}=T_eG$ be an element of the Lie algebra of $G$. We saw
in section \ref{alan:lie} that $A$ generates a one-parameter flow
$\sigma_t(g)$ through $g$ in $G$. A slight modification of this
construction shows that $A$ will generate a flow in $P$ along the
fibre at each point of $M$ by the right action of $G$ on $P$:
 \be
 \sigma_t(p) = R_{\exp(tA)}p = p\, \exp(tA).
 \ee
Note that $\pi(p)=\pi(\sigma_t(p))$, so that indeed vectors tangent
to the curves are elements of $V_pP$. We now define a map
$\mathfrak{g}\rightarrow V_pP$ which maps $A$ to the vector tangent
to $\sigma_t(p)$ for $t=0$, which we will call (with slight abuse of
notation with respect to equation (\ref{alan:flow_eq})) $X_A\in
V_pP$. The equivalent of the flow equation (\ref{alan:formalflow})
now becomes
 \be
 X_A(f(p))=\left. \frac{d}{dt}f(\sigma_t(p))\right\vert_{t=0}.
 \label{alan:fund_vectorfield}
 \ee
$X_A$ is called the fundamental vector field associated with $A$.
The fundamental vector fields associated to a basis of the Lie
algebra form a basis of the vertical subspace. A connection one-form
is now defined as follows.

\begin{defi}
\label{alan:con_one-form} A {\bf connection one-form} $\omega\in \Lambda P
\otimes \mathfrak{g}$ (where $\Lambda P \equiv \Gamma(P,T^*P)$) is a Lie algebra valued one-form defined by a
projection of the tangent space $T_pP$ onto the vertical subspace
$V_pP$ satisfying
\renewcommand{\labelenumi}{(\roman{enumi})}
\begin{enumerate}
\item $\omega(X_A)=A$ for every $A\in \mathfrak{g}$;
\item $R_g^*\omega = Ad_{g^{-1}}\omega$ for every $g\in G$.
\end{enumerate}
\end{defi}

More concretely, (i) means that $\omega$ acts as a Maurer-Cartan
form on the vertical subspace and (ii) means that for $X\in T_pP$,
 \be
 R_g^{~*}\omega\vert_p(X) = \omega\vert_{pg}(R_{g*}X) =
 g^{-1}\omega(X)g.
 \ee
The horizontal subspace is then defined as the
set
 \be
 H_pP = \{X \in T_pP\vert\, \omega(X)=0\}.
 \ee
When defined in this way, $H_pP$ still satisfies the equivariance
condition. To see this, take $X \in H_pP$ and construct $R_{g*}X\in
T_{pg}P$. This is an element of $H_{pg}$ because
 \be
 \omega(R_{g*}X) = R_g^{~*}\omega (X) = g^{-1}\omega(X)g=0.
 \ee
So both definitions of a connection are equivalent. This connection
is defined over all of $P$. To connect to physics, we have to relate
this to a one-form in the base $M$. It turns out that this can only
be done locally (when $P$ is non-trivial).

\begin{defi}
Let $\{U_i\}$ be an open covering of $M$. Choose a local section
$s_i$ on $U_i$
 \be
 s_i:U_i \rightarrow \pi^{-1}(U_i).
 \ee
The local connection one-form or {\bf gauge potential} is now defined as
 \be
 A_i \equiv s_i^{~*}\omega \in \Gamma(U_i,T^*M,)\otimes
 \mathfrak{g} = \Lambda U_i \otimes
 \mathfrak{g} \label{alan:gaugecon}
 \ee
\end{defi}

Also the converse is true.

\begin{theo}
Given a local connection one-form $A_i$ and a section $s_i$ on an
open subset $U_i \subset M$, there is a unique connection one-form
$\omega \in \pi^{-1}(U_i)$ such that $A_i = s_i^{~*}\omega$.
\end{theo}

We will prove this theorem rather explicitly, since this will give
better insight into the emergence of a gauge potential on $M$ from
the connection one-form on $P$. First of all, we introduce the
notion of a canonical local trivialization with respect to a
section. Given a section $s_i$ on $U_i$ and a $p\in \pi^{-1}(U_i)$,
there always exists a $g_i\in G$ such that $p = s_i(x)g_i$, where $x
= \pi(p)$. This means that we can define a local trivialization by
 \be
 \phi^{-1}: \pi^{-1}(U_i) \rightarrow U_i \times G: p \mapsto
 (x,\,g_i).
 \ee
This means that the section itself is represented as
$s_i(x)=(x,\,e)$. On an overlap $U_i \cap U_j \neq \emptyset$ two
sections are related by
 \be
 s_i(x) &=& \phi_i(x,\,e) = \phi_j(x,\,g_{ji}(x)e)=
 \phi_j(x,\,g_{ji}(x))\nonumber\\
 &=& \phi_j(x,\,e)g_{ji}(x) = s_j(x) g_{ji}(x).
 \ee
First we proof that an $\omega$ exists, then we will sketch a proof
of its uniqueness.

\vspace{.4cm}

\noindent {\bf Proof (existence):}

\noindent Given a section $s_i$ and a gauge potential $A_i$ on
$U_i$, we propose the following form of the connection one-form:
 \be
 \omega\vert_{U_i} = g_i^{-1}\pi^* A_ig_i + g_i^{-1}d_P g_i\,,
 \label{alan:from_A_to_w}
 \ee
where $d_P$ is the exterior derivative on $P$ and $g_i$ is the group
element which appears in the definition of the canonical local
trivialization with respect to $s_i$.
\renewcommand{\labelenumi}{(\roman{enumi})}
\begin{enumerate}
\item First of all, we have to show that pulling back
(\ref{alan:from_A_to_w}) with the section results in $A_i$. Note
that since $\pi \circ s_i = \mbox{Id}_{U_i}$, we have that $\pi_*
s_{i*} = \mbox{Id}_{TU_i}$ and that $g_i=e$ on $s_i$. For a $X\in
T_xM$ we have
 \be
 s_i^{~*}\omega\vert_{U_i}(X) &=& \omega(s_{i*}X) = \pi^*A_i(s_{i*}X)
 + d_P e(s_{i*}X)\nonumber\\
 &=& A_i(\pi_* s_{i*}X) = A_i(X).
 \ee
\item Now we need to establish that (\ref{alan:from_A_to_w})
satisfies the conditions from definition \ref{alan:con_one-form}. A
fundamental vector field $V_A$ satisfies $\pi_* X_A = 0$ so that
only the second term from (\ref{alan:from_A_to_w}) contributes. We
need to evaluate this in the sense of a Maurer-Cartan form as
discussed in subsection (\ref{alan:lie}),
 \be
 w\vert_{U_i}(X_A) &=& g_i^{-1}d_P g_i(X_A) =
 g_i^{-1}X_A(g_i) = g_i^{-1}X_A\vert_{g_i}\nonumber\\
 &=& g_i^{-1}(p)\left.
 \frac{d}{dt}g_i(\sigma_t(p))\right\vert_{t=0}
 = g_i^{-1}(p)\left.
 \frac{d}{dt}g_i(p\, \exp(tA))\right\vert_{t=0} \nonumber\\
 &=& g_i^{-1}(p)g_i(p)\left.
 \frac{d}{dt}\exp(tA)\right\vert_{t=0} = \left.
 \frac{d \sigma_t(e)}{dt}\right\vert_{t=0} = A,
 \ee
where we used both the definition of the fundamental vector field
(\ref{alan:fund_vectorfield}) and the flow equation
(\ref{alan:flow_eq}). This proves the first condition for being a
connection. To prove the second condition, take an $X\in T_pP$. Note
next that $g_i(p\,h)=g_i(p)h$ and that since $\pi\circ R_h=\pi$, we
have that $\pi_* R_{h*} = \pi_*$. We find
 \be
 R_h^{~*}\omega(X) &=& \omega(R_{h*}X) = h^{-1}g_i^{-1}A_i(\pi_* X)g_i
 h + h^{-1}g_i^{-1} d_P g_i(X)h \nonumber\\
 &=& h^{-1}\omega(X)h = Ad_{h^{-1}}\omega(X).
 \ee
\end{enumerate}
\begin{flushright}
$\Box$
\end{flushright}

We will now sketch the proof of the uniqueness of the connection
one-form. For this we need to see what happens on overlaps $U_i\cap
U_j \neq \emptyset$.

\vspace{.4cm}

\noindent {\bf Proof (uniqueness):}

\noindent The two definitions of the connection on each
patch have to agree on the intersection, $\omega\vert_{U_i} =
\omega\vert_{U_j}$ on $U_i\cap
U_j \neq \emptyset$. Writing this out, we find the following condition:
 \be
 g_i^{-1}\pi^* A_ig_i + g_i^{-1}d_P g_i = g_j^{-1}\pi^* A_jg_j + g_j^{-1}d_P g_j
 \ee
Noting that on the intersection we have $g_j = g_{ji}g_i$ ($s_j =
s_ig_{ij}$), a small calculation shows that
 \be
 \pi^*A_j = g_{ij}^{-1}\pi^* A_ig_{ij} + g_{ij}^{-1}d_P g_{ij}.
 \ee
We can use either one of the sections $s_{i,j}$ on $U_i\cap U_j$ to
pull this back to a local statement (note that $s_i^*\pi^* =
\mbox{Id}_M$ and that the pull-back and the exterior derivative
commute),
 \be
 A_j = g_{ij}^{-1}A_ig_{ij} + g_{ij}^{-1}d g_{ij}.
 \ee
So we see that both definitions of $\omega$ agree on $U_i\cap
U_j \neq \emptyset$ if the local gauge potentials are related in the above way.
\begin{flushright}
$\Box$
\end{flushright}

This is the important point we wanted to reach. Note that the
connection one-form $\omega$ on $P$ is defined globally; it contains
global information on the non-triviality of $P$. The gauge
potentials $\{A_i\}$ are defined locally and we have just seen that
if the fibres over two intersecting open sets on the base space have
to be identified in a non-trivial way, the two gauge potentials
defined on the overlap are necessarily different. This means that on
a non-trivial bundle one local gauge potential has no global
information, only the collection of all locally defined gauge
potentials knows about the global topology. This means that gauge
freedom is sometimes not just a matter of choice, but more of
necessity!

Of course also in this language gauge freedom is a reflection of
choice. Say that on an open set $U$, two sections are related by
$s'(x) = s(x)g(x)$. We can choose either section to define a local
gauge potential and almost the same reasoning as above shows that
both are related as follows
 \be
 A'(x) = g(x)^{-1}A(x)g(x) + g(x)^{-1}d g(x). \label{alan:gaugetransfo}
 \ee
We see that in the bundle language, gauge freedom is equivalent to
the freedom to choose local coordinates on a principal bundle!

\subsection{Curvature of a connection}

A very important notion is of course the curvature of a connection.
To introduce this, we first define the concept of covariant
derivative on a principal bundle.

\begin{defi}
Consider a Lie algebra valued $p$-form, $\alpha \in \Lambda^nP
\otimes \mathfrak{g}$. This can be decomposed as $\alpha = \alpha^a
\otimes T_a$, where $\alpha^a$ is an ordinary $p$-form and $T_a$ is
a basis of $\mathfrak{g}$. Let $X_1,...,X_{p+1}\in T_pP$ be $p+1$
tangent vectors on $P$. The {\bf exterior covariant derivative} of
$\alpha$ is defined by
 \be
 D\alpha(X_1,...,X_{p+1}) = d_P\alpha (X_1^H,...,X_{p+1}^H),
 \ee
where $X_i^H\in H_pP$ is the horizontal component of $X_i\in T_pP$ and $d_P\alpha = (d_P\alpha^a) \otimes T_a$.
\end{defi}

The curvature is now readily defined.

\begin{defi}
The {\bf curvature 2-form} $\Omega$ of a connection $\omega$ on $P$ is defined as
 \be
 \Omega = D\omega \in \Lambda^2 P \otimes \mathfrak{g}
 \ee
\end{defi}

At every point $p\in P$, the horizontal vectors define a subspace of
$T_pP$. The assignment of such a subspace at every point of $P$ is
called a distribution. Since in this case the distribution is
defined by the equation $\omega = 0$, the Frobenius condition for
integrability of the submanifold of $P$ tangent to this distribution
is exactly the vanishing of $d\omega$ along the distribution, that
is, the vanishing of the curvature. As such, the curvature is an
obstruction to finding a submanifold of $P$ that is `completely
horizontal'.

\begin{theo}
The curvature 2-form has the property
 \be
 R_g^{~*}\Omega = Ad_{g^{-1}}\Omega = g^{-1}\Omega g,
 \ee
where $g$ is a constant element of $G$.
\end{theo}

\vspace{.4cm}

\noindent {\bf Proof:}

\noindent Recall that, because of the equivariance property of
horizontal subspaces
 \be
 &&(R_{g*}X)^H = R_{g*}X^H;\nonumber\\
 &&R_g^{~*}\omega = Ad_{g^{-1}}\omega = g^{-1}\omega g. \nonumber
 \ee
Also recall that pull-backs and exterior derivatives commute, $d_P
R_g^{~*} = R_g^{~*}d_P$ and because $g$ is constant, $d_Pg=0$. For
$X$, $Y \in T_pP$ we then have
 \be
 R_g^{~*}\Omega(X,Y) &=& \Omega(R_{g*}X,R_{g*}Y) = d_P\omega(R_{g*}X^H,R_{g*}Y^H)\nonumber\\
 &=& R_g^{~*}d_P\omega(X^H,Y^H) = d_P R_g^{~*}\omega(X^H,Y^H)\nonumber\\
 &=& d_P (g^{-1}\omega g)(X^H,Y^H) = g^{-1}d_P \omega(X^H,Y^H)g\nonumber\\
 &=& g^{-1}\Omega(X,Y)g
 \ee
\begin{flushright}
$\Box$
\end{flushright}

Let $\alpha\in\Lambda^p\otimes\mathfrak{g}$ and
$\beta\in\Lambda^q\otimes\mathfrak{g}$ be two Lie algebra valued
forms. The Lie bracket (commutator) between the two is defined as
 \be
 [\alpha,\,\beta] &\equiv& \alpha\wedge\beta -(-)^{pq} \beta\wedge\alpha \nonumber\\
 &=& T_a T_b \alpha^a\wedge\beta^b -(-)^{pq} T_b T_a \beta^b\wedge\alpha^a \nonumber\\
 &=& [T_a,T_b]\alpha^a\wedge\beta^b = f_{ab}^{~~c}T_c\, \alpha^a\wedge\beta^b.
 \ee
Note that for odd $p$ this means that
 \be
 [\alpha,\alpha] = 2\alpha\wedge\alpha \neq 0.
 \ee
For even $p$, $[\alpha,\alpha]=0$.

Let us now proof the following important theorem:

\begin{theo}
$\Omega$ and $\omega$ satisfy the Cartan structure equations ($X$, $Y\in T_pP$)
 \be
 \Omega(X,Y)=d_P\omega(X,Y)+ [\omega(X),\omega(Y)],\label{alan:cartan}
 \ee
or
 \be
 \Omega = d_P\omega + \omega\wedge\omega = d_P\omega + \frac 1 2 [\omega,\omega].
 \ee
\end{theo}

To see the relation between the two forms of the theorem note that
 \be
 [\omega,\omega](X,Y) &=& [T_a,T_b]\omega^a\wedge\omega^b (X,Y) \nonumber\\
 &=& [T_a,T_b](\omega^a(X)\omega^b(Y)-\omega^a(Y)\omega^b(X))\nonumber\\
 &=& [\omega(X),\omega(Y)] - [\omega(Y),\omega(X)] = 2[\omega(X),\omega(Y)].\nonumber
 \ee

\vspace{.4cm}

\noindent {\bf Proof:}

\noindent We consider three cases separately
\renewcommand{\labelenumi}{(\roman{enumi})}
\begin{enumerate}
\item Let $X$, $Y\in H_pP$. Then by definition $\omega(X)=\omega(Y)=0$. Then (\ref{alan:cartan})
follows trivially since by definition
 \be
 \Omega(X,Y)=d_P\omega(X^H,Y^H)=d_P\omega(X,Y).\nonumber
 \ee
\item Let $X\in H_pP$ and $Y\in V_pP$. Since $Y^H=0$, by definition $\Omega(X,Y)=0$.
We still have that $\omega(X)=0$, so we still have to prove that $d_pP\omega(X,Y)=0$.
To do this, we use the following identity:
 \be
 d_P\omega(X,Y) &=& X\omega(Y) - Y\omega(X) -\omega([X,Y]) \nonumber\\
 &=& X\omega(Y) -\omega([X,Y]). \nonumber
 \ee
Since $Y\in V_pP$, it is a fundamental vector field\footnote{Or a linear combination
of fundamental vector fields, in which case the result follows by linearity} for some
$V\in\mathfrak{g}$. This means that $\omega(Y)=V$ is a constant, so $X\omega(Y) = XV = 0$.
One can show that $[X,Y]\in H_pP$, so that also $\omega([X,Y])=0$.
\item Let $X$, $Y\in V_pP$. Then again $\Omega(X,Y)=0$ and this time we have
 \be
 d_P\omega(X,Y) = -\omega([X,Y]). \nonumber
 \ee
So we still have to prove $\omega([X,Y]) = [\omega(X),\omega(Y)]$.
Since also $[X,Y]\in V_pP$, every vector $X=X_B$, $Y=X_C$ and
$[X,Y]=X_A$ are fundamental vector fields associated to Lie algebra
elements $B$, $C$ and $A$ respectively. One can prove that
necessarily $A=[B,C]$, which completes the proof.
\end{enumerate}
\begin{flushright}
$\Box$
\end{flushright}

We will now again use a section to pull back this globally defined
object on $P$ to a local object defined on a patch on $M$.

\begin{defi}
Given a section $s_i$ on $U_i$, the local (Yang-Mills) {\bf field strength} is defined by
 \be
 F_i = s_i^{~*}\Omega \in \Lambda^2 U_i \otimes\mathfrak{g}.
 \ee
\end{defi}

The relation with the gauge potential is now easily obtained
 \be
 F_i &=& s_i^{~*}d_P\omega + s_i^{~*}(\omega\wedge\omega) = d (s_i^{~*}\omega)
  + s_i^{~*}\omega\wedge s_i^{~*}\omega \nonumber\\
 &=& dA_i + A_i\wedge A_i.
 \ee
Writing $A=A_a dx^a$ and $F=\frac 1 2 F_{ab}dx^a\wedge dx^b$ (we
dropped the subscript $i$ for convenience), we find the usual
expression,
 \be
 F_{ab} = \partial_a A_b - \partial_b A_a + [A_a,A_b].
 \ee
The effect of a coordinate change on the field strength 2-form can
be deduced (as usual) from the transformation properties of the
gauge potential 1-form (\ref{alan:gaugetransfo}). More specifically,
if two sections are related by $s'(x)=s(x)g(x)$, the corresponding
field strengths are related by
 \be
 F'(x) = g(x)^{-1}F(x)g(x).
 \ee

We will need an important identity involving the curvature. Since,
$\omega(X)=0$ for $X\in H_pP$, we find that for $X$, $Y$, $Z\in
T_pP$
 \be
 D\Omega(X,Y,Z) = d_P\Omega(X^H,Y^H,Z^H) = (d_P\omega \wedge \omega -
 \omega \wedge d_P\omega)(X^H,Y^H,Z^H) = 0. \nonumber
 \ee
This proves the {\bf Bianchi identity}
 \be
 D\Omega = 0.
 \ee
To find the local form of this identity we use a section $s_i$ to
pull back the relation
 \be
 d_P\Omega = d_P\omega \wedge \omega -
 \omega \wedge d_P\omega.
 \ee
This results in
 \be
 dF_i &=& d s_i^{~*}\Omega = s_i^{~*}d_P\Omega = s_i^{~*}(d_P\omega \wedge \omega -
 \omega \wedge d_P\omega)\nonumber\\
 &=& d s_i^{~*}\omega \wedge s_i^{~*}\omega - s_i^{~*}\omega \wedge
 ds_i^{~*}\omega = dA_i\wedge A_i - A_i \wedge dA_i \nonumber\\
 &=& F_i \wedge A_i - A_i \wedge F_i = -[A_i,F_i].\nonumber
 \ee
So we find the local identity
 \be
 {\cal D}_iF_i = dF_i + [A_i,F_i] = 0,\label{alan:local_Bianchi}
 \ee
where we defined the {\bf covariant derivative}
 \be
 {\cal D}_i = d + [A_i,~~].
 \ee

\section{The topology of principal bundles}
\label{alan:topology}

We will now discuss some aspects of the topology of gauge bundles.
As we have seen, pure Yang-Mills theory can be described using only
principal bundles, so we restrict ourselves to a discussion of the
topology of principal bundles. Again, this section is mainly
influenced by \cite{Nakahara}, \cite{Bertlmann} and \cite{Frankel}.
For more advanced treatments and different perspectives, we
recommend \cite{Steenrod}, \cite{Milnor} and \cite{Nash}.

\subsection{Aspects of homotopy theory}
\label{alan:homotopy}

We start by making some simple remarks about the classification of
topological spaces. Usually in topology, two spaces are considered
equivalent if they can continuously be deformed into each other. In
other words, they are considered topologically the same, if there
exists a homeomorphism between them (for differentiable manifolds
this would have to be diffeomorphism). Classifying spaces up to
homeomorphism is a difficult thing to do. The idea is to find as
much topological invariants (in general, numbers that do not depend
on continuous parameters) of a type of space as possible. Finding a
full set of invariants that completely classifies a space is rather
difficult. The converse is however easily stated:

\begin{quote}
If two topological spaces have different topological invariants,
they are not homeomorphic, hence not topologically equivalent.
\end{quote}

Since classification up to homeomorphism is such a difficult task,
one can try to answer somewhat easier problems. For instance, one
can try to classify spaces up to homotopy. Two spaces are said to be
homotopic to each other if one can be mapped to the other in a
continuous way, but this map need not have an inverse. For instance
a circle and a cylinder are homotopic (one can continuously shrink
the cylinder until its length disappears), but clearly not
homeomorphic. In an intuitive sense, homotopic equivalence occurs
when in the process of deforming one space to another, a part of the
space is `lost', so that it becomes impossible to define the reverse
process.

For the moment, we are interested in homotopic equivalence classes
of loops on a differentiable manifold $M$. It turns out that these
reveal a very interesting group structure. We will only sketch this
construction and state the results that we need for the remainder of
these notes. By a (based) loop we will mean a map $\alpha:
I=[0,1]\rightarrow M: t \mapsto \alpha(t)$, such that
$\alpha(0)=\alpha(1)=x\in M$, the base point of the loop. It turns
out that not these loops by themselves exhibit a group structure,
but rather their equivalence classes under homopoty.

\begin{defi}
Two loops $\alpha$ and $\beta$ based at the point $x\in M$ are
{\bf homotopic} to each other if there exists a continuous map
 \be
 H: I\times I \rightarrow: M: (s,t)\mapsto H(s,t)
 \ee
such that
\begin{center}
\begin{tabular}{ll}
 $H(s,0) = \alpha(s)$ and $H(s,1) = \beta(s)$,& $\forall s\in I$,\\
 $H(0,t) = H(1,t) = x$, & $\forall t\in I$.
\end{tabular}
\end{center}
$H(s,t)$ is called the {\bf homotopy} between $\alpha$ and $\beta$.
\end{defi}

One can show that this is an equivalence relation $\sim$ (reflexive,
symmetric, transitive) between based loops. We will denote the
equivalence class or homotopy class by $[\alpha]$. So we have that
 \be
 \alpha \sim \beta \Rightarrow [\alpha] = [\beta].
 \ee
In other words, two loops are considered the same if one can
continuously deform one into the other. One can now define the
`product' $\alpha \circ \beta$ of two loops as the result of first
going through $\alpha$ and then through $\beta$. The inverse
$\alpha^{-1}$ of a loop is then just going through the loop $\alpha$
in reverse order and the unit element is the constant loop
$\alpha(t)=x$, $\forall t\in I$. It is clear that $\alpha \circ
\alpha^{-1}$ are not equal, but homotopic to the identity. This is
why only the homotopy classes of loops exhibit a group structure.

\begin{defi}
The group formed by the homotopy classes of loops based at $x$ on a
manifold $M$ is called the {\bf fundamental group} or first homotopy
group $\Pi_1(M,x)$.
\end{defi}

One can show that, if the manifold is arc-wise connected, the
fundamental groups at different points are isomorphic. In that case
we just refer to the fundamental group of the manifold $\Pi_1(M)$.
If two manifolds are homotopic (of the same homotopy type) one can
show that their fundamental groups are isomorphic. Since homotopy is
a weaker than homeomorphy, if the fundamental group is invariant
under homotopy, it must certainly be under homeomorphy. So we arrive
at the following conclusion.

\begin{theo}
The fundamental group is invariant under homeomorphisms and hence is
a topological invariant.
\end{theo}

As an example, let us look at $\Pi_1(S^1)=\Pi_1(U(1))$. This
basically means that we want to classify maps from one circle to
another. Intuitively, we know that one circle can wind the other an
integer $n$ times. This is accomplished by functions of the form
 \be
 g_{n,a}: I \rightarrow S^1: t \mapsto g_{n,a}(t) = e^{i(nt+a)}, \quad a\in \mathbb{R}.
 \ee
It's easily shown that two maps $g_{n,a}$ and $g_{m,b}$ are
homotopic to each other for any $a,\,b\in \mathbb{R}$ if $n=m$, but
that for $n \neq m$ they are homotopically distinct. This means that
we have homotopy classes
 \be
 [n] \equiv [g_{n}],
 \ee
where $g_{n} \equiv g_{n,0} = e^{int}$ is a good representative for
each equivalence class. In this easy case one calls the integer $n$
the degree or winding number of the map and we find that homotopy
classes are characterized by their winding number. This winding
number can be represented by the integral
 \be
 n = \frac{1}{2\pi i}\int_0^{2\pi} dt\, g_n(t)^{-1}\frac{d}{dt}g_n
 (t). \label{alan:degcircle}
 \ee
From this it is clear that if $f_1(t)$ represents a map of winding
number one, a map of winding number $n$ is represented by $f_n(t) =
f_1(t)^n$. It is also clear that the product of $[g_n]$ and $[g_m]$
is nothing but $[g_{n+m}]$, which means that the fundamental group
is isomorphic to the additive group $\mathbb{Z}$. So we get the well
known result $\Pi_1(S^1)=\mathbb{Z}$. In general $\Pi_1(M)$ can be
non-Abelian.

One can also define the higher homotopy groups by looking at
homotopy classes of maps from higher dimensional spheres to a
manifold. More concretely, one looks at maps $\alpha: I^n
\rightarrow M$, such that the entire boundary of the $n$-dimensional
cube, $\partial I^n$ maps to a single point $x\in M$. Going through
the same procedure as for $\Pi_1(M)$ one arrives at the higher
homotopy groups $\Pi_n(M)$, which are always Abelian. In general
these homotopy groups are quite hard to calculate. However, in the
examples we will be studying, the maps we want to classify are
always between two spaces of the same dimension. In that case, there
is the notion of the (Brouwer) degree of a map, which is somewhat
easier to handle. In the end, it will give us the same information
as the related homotopy groups would provide.

\begin{defi}
Consider a map $\phi: M\rightarrow N$, where dim$M=$ dim$N=n$ and
let $\Omega$ be a normalized volume form on $N$. The {\bf Brouwer
degree} of this map is defined as
 \be
 \mbox{deg}(\phi)= \int_M \phi^*\Omega,\quad \int_N \Omega = 1
 \label{alan:degree}
 \ee
\end{defi}

This definition does not depend on the volume form chosen, since the
difference of two normalized volume forms has to be exact (its
integral over $N$ has to vanish) and the pull-back commutes with the
exterior derivative. In addition one can show that the degree is an
integer and hence has to be a topological invariant. We will show
below that the degree we defined in (\ref{alan:degcircle}) for the
circle can interpreted exactly in this way.

We would like to study the topology of the simplest non-trivial
bundles. Since, as we already mentioned, a bundle over a
contractible space is always trivial, bundles over $\mathbb{R}^n$
are always trivial. The next simplest thing to study are bundles
over $n$-spheres $S^n$ and since these bundles are also very
relevant and interesting for physics, we will mainly focus on these.
One can always cover an $n$-sphere by two patches, say the north and
the south hemisphere. The intersection of these two patches is
homotopic to the equator, an $(n-1)$-sphere. This means that to
classify a principal $G$-bundle over $S^n$, one would have to
classify the transition functions on $S^{n-1}$, that is all maps
from $S^{n-1}$ to $G$. As we already discussed, a very interesting
object to study in this regard is the homotopy group $\Pi_{n-1}(G)$.
Later on we will look at $U(1)$-bundles over $S^2$ (Dirac monopoles)
and $SU(2)$-bundles over $S^4$ (instantons). This requires knowledge
of the groups $\Pi_1(U(1))$ and $\Pi_3(SU(2))$, respectively.
$\Pi_1(U(1))=\mathbb{Z}$ was already considered in the previous
example, so let's now focus on $\Pi_3(SU(2))$.

We will again use the fact that homotopy classes of maps from $S^3$
to $S^3$ are characterized by their topological degree. First of
all, we have to find a well defined volume form on $SU(2)$. For a
general compact (matrix) group manifold, this is done as follows. At
every point $g\in G$, one can define the left invariant
$\mathfrak{g}$-valued Maurer-Cartan form $\Theta = g^{-1}dg$, as
discussed in subsection (\ref{alan:lie}). From this one can define a
well-defined bi-invariant (left- and right-invariant) volume form on
$G$. For instance, for $SU(2)$ a normalized volume form is given by
 \be
 \Omega = \frac{1}{24\pi^2}\Tr \left( g^{-1}dg\wedge g^{-1}dg\wedge g^{-1}dg
 \right),
 \ee
where,
 \be
 g = c_0 1_2 + c_i \tau_i \, , \qquad c_0^2 + c_i c_i = 1,
 \ee
and $\tau_i$, $i\in\{1,2,3\}$ are the Pauli matrices. Let (with a
slight abuse of notation) $g:S^3\rightarrow SU(2): x \mapsto g(x)$.
Then the degree of this map, according to equation
(\ref{alan:degree}), is given by
 \be
 \mbox{deg}(g) = \frac{1}{24\pi^2} \int_{S^3} \Tr \left( g^{-1}dg\wedge g^{-1}dg\wedge g^{-1}dg
 \right), \label{alan:deg3sphere}
 \ee
where the integrand should now be interpreted as the pull-back
$g^*\Omega$. In other words, $dg$ should now be interpreted as
$\partial_i g dx^i$. Again, equation (\ref{alan:deg3sphere}) will
always give an integer $n$ and since this degree fully characterizes
elements of $\Pi_3(SU(2))$, we find that $\Pi_3(SU(2))=\mathbb{Z}$.

In the case $G=U(1)$, the Maurer-Cartan form itself is a
bi-invariant volume form, so we can take
 \be
 \Omega = \frac{1}{2\pi i}\, g^{-1}dg\,,\qquad g\in U(1)
 \ee
For the map $g_n:S^1\rightarrow U(1): t\mapsto g_n(t)$ we considered
above we get
 \be
 g_n^{~*}\Omega = \frac{1}{2\pi i}\,
 g_n(t)^{-1}\frac{dg_n(t)}{dt}\,dt\,,
 \ee
so that the integer $n$ we defined for the circle above, can
rightfully be called the degree of the map $g_n$,
$\mbox{deg}(g_n)=n$.

\subsection{Characteristic classes}

Besides Homotopy there are of course many different ways to
construct topological invariants. An important example are groups
generated by (co)homology classes of a manifold. We will now focus
on certain integer cohomology classes constructed from polynomials
in the field strength of a bundle, called characteristic classes.
First we define invariant polynomials.

\begin{defi}
Let $\mathfrak{g}$ be the Lie algebra of some $G$. A totally
symmetric and $n$-linear polynomial
 \be
 P(X_1,...,X_i,...,X_j,...,X_n)= P(X_1,...,X_j,...,X_i,...,X_n),\quad \forall\, i,j,
 \ee
where $X_i$, $i\in\{1,...,n\}$ are elements of $\mathfrak{g}$, is
called a {\bf symmetric invariant} (or characteristic) {\bf
polynomial} if
 \be
 P(g^{-1}X_1g,...,g^{-1}X_ng) = P(X_1,...,X_n),\quad g\in G.\label{alan:invariant}
 \ee
\end{defi}

An immediate consequence of this definition is that (take $g$ close
to the identity, $g=1+tY$, and expand (\ref{alan:invariant}) to
first order in $t$),
 \be
 \sum_{i=1}^n P(X_1,...,X_{i-1},[Y,X_i],X_{i+1},...,X_n)=0. \label{alan:linear_inv}
 \ee
This will be of use to us later.

\begin{defi}
An {\bf invariant polynomial} of degree $n$ is defined as a
symmetric invariant polynomial with all its entries equal,
 \be
 P_n(X) \equiv P(\mathop{\underbrace{X,...,X}}_n)\ \equiv P(X^n)
 \ee
\end{defi}

Now we want to extend this definition to $\mathfrak{g}$-valued
differential forms on a manifold. A $\mathfrak{g}$-valued $p$-form,
$\alpha_i$, we write as (no sum over $i$) $\alpha_i = \eta_i X_i$,
where $\eta_i$ is an ordinary $p$-form and $X_i$ again an element of
$\mathfrak{g}$. We then extend the previous definition as follows:

\begin{defi}
An invariant polynomial for $\mathfrak{g}$-valued forms is defined
as
 \be
 P(\alpha_1,...,\alpha_n) = P(X_1,...,X_n)\, \eta_1 \wedge...\wedge \eta_n.
 \ee
The diagonal combination is again called an invariant polynomial of
degree $n$,
 \be
 P_n(\alpha)=P(\alpha^n)= P(X^n)\, \eta\wedge...\wedge\eta.
 \ee
\end{defi}

Let $\beta$ be a $\mathfrak{g}$-valued 1-form. From
(\ref{alan:linear_inv}) we find that
 \be
 \sum_{i=1}^n (-)^{p_1 + ... + p_{i-1}}P(\alpha_1,...,\alpha_{i-1},
 [\beta,\alpha_i],\alpha_{i+1},...,\alpha_n) = 0,
 \ee
where $p_i$ is the degree of $\alpha_i$ and the minus signs arise
from pulling the 1-form to the front each time. Equally easy to
compute is
 \be
 dP(\alpha_1,...,\alpha_n) = \sum_{i=1}^n (-)^{p_1 + ... + p_{i-1}}
 P(\alpha_1,...,\alpha_{i-1},d\alpha_i,\alpha_{i+1},...,\alpha_n).
 \ee
Adding both equations for the specific case $\beta = A$, where $A$
is the local gauge potential associated to a connection on the
bundle (we drop the index referring to the specific local patch for
convenience) and recalling the expression for the covariant
derivative ${\cal D}=d+[A,~~]$, we find the important expression
 \be
 dP(\alpha_1,...,\alpha_n) = \sum_{i=1}^n (-)^{p_1 + ... + p_{i-1}}
 P(\alpha_1,...,\alpha_{i-1},{\cal D}\alpha_i,\alpha_{i+1},...,\alpha_n).\label{alan:technical_result}
 \ee

The objects we want to study are invariant polynomials in the field
strength 2-form $F$, $P_n(F)$, because these turn out to have very
interesting properties. We will work with a connection on a
principal bundle, although the following results equally hold for
its associated vector bundle. We are now ready to prove the
following important theorem:

\begin{theo}
Let $P_n(F)$ be an invariant polynomial, then
\renewcommand{\labelenumi}{(\roman{enumi})}
\begin{enumerate}
\item $P_n(F)$ is closed, $dP_n(F)=0$.
\item Let $F$ and $F'$ be local curvature 2-forms corresponding
to two different connections on the same bundle. Then the difference $P_n(F)-P_n(F')$ is exact.
\end{enumerate}
\end{theo}

Note that, since $P_n(F)$ is closed, it can always locally be
written as the $d$ of something (Poincar\'e's lemma). The important
point about this theorem is that the difference $P_n(F)-P_n(F') =
dQ$ is exact in a global sense (which is of course the meaning of
exact), meaning that we have to prove that $Q$ is globally defined!
(a point which is usually ignored in physics textbooks)

\vspace{.4cm}

\noindent {\bf Proof:}

\renewcommand{\labelenumi}{(\roman{enumi})}
\begin{enumerate}
\item The first part of the theorem follows immediately from
(\ref{alan:technical_result}), because of the Bianchi identity
${\cal D}F=0$, see (\ref{alan:local_Bianchi}).
\item To prove the second part, consider two 1-form gauge potentials
$A$ and $A'$, both referring to the same system of local trivializations,
and their respective 2-form field strengths $F$ and $F'$. We define the homotopic
connection\footnote{We are a bit sloppy here, because we should define this homotopy
locally on a patch, where it is clear that this can be done. However, since both
connections are defined on the same bundle (same set of transition functions) this
turns out to be possible globally.}
 \be
 A_t = A + t\theta, \quad \theta = A' - A,
 \ee
so that $A_0 = A$ and $A_1 = A'$, and its field strength
 \be
 F_t &=& dA_t + A_t \wedge A_t = F + t(d\theta + A\wedge\theta + \theta\wedge A) + t^2
 \theta\wedge\theta \nonumber\\
 &=& F + t{\cal D}\theta + t^2\theta\wedge\theta,
 \ee
where ${\cal D}= d + [A,~~]$ (note the sign convention in the
definition of the commutator). We now differentiate $F_t$ with
respect to $t$,
 \be
 \frac{d}{dt}F_t &=& {\cal D}\theta + 2t\theta\wedge\theta = d\theta +
 A_t\wedge\theta + \theta\wedge A_t = {\cal D}_t\theta
 \ee
with the obvious notation ${\cal D}_t = d + [A_t,~~]$. Considering
then the invariant polynomial $P_n(F_t)$, we get
 \be
 \frac{d}{dt} P_n(F_t) = nP(\frac{d}{dt}F_t,\mathop{\underbrace{F_t,...,F_t}}_{n-1})
 = nP({\cal D}_t\theta,F_t^{n-1}).
 \ee
From equation (\ref{alan:technical_result})  and ${\cal D}_tF_t=0$,
we know that $dP(\theta,F_t^{n-1})=P({\cal D}_t\theta,F_t^{n-1})$,
so that we find that
 \be
 \frac{d}{dt} P_n(F_t) = n dP(\theta,F_t^{n-1}).
 \ee
Integrating this from $t=0$ to $t=1$, we find
 \be
 P_n(F')-P_n(F) = dQ_{2n-1}(A',A),\label{alan:transgression}
 \ee
where we defined the {\bf transgression} $Q_{2n-1}(A',A)$ as
 \be
 Q_{2n-1}(A',A) = n\int_0^1 dt\, P(A'-A,F_t^{n-1}).
 \ee
Note that $Q_{2n-1}(A',A)$ is indeed a gauge invariant and hence
globally defined object, since under a gauge transformation (the
inhomogeneous term cancels) $\theta' = g^{-1}\theta g$ and $P$ is
invariant.
\end{enumerate}
\begin{flushright}
$\Box$
\end{flushright}

Equation (\ref{alan:transgression}) is called a transgression
formula and is quite important in the study of anomalies. We use it
here to define the {\bf Chern-Simons form}. Say that one can define
a trivial connection $A'=0$ on a bundle. This means that either the
bundle is trivial or that we are working on a local patch. We know
that since $P_n(F)$ is closed, it is locally exact; it can locally
be written as the $d$ of a Chern-Simons form. The transgression
formula provides a means for calculating this Chern-Simons form.
Indeed, from (\ref{alan:transgression}) we find
 \be
 P_n(F) = dQ_{2n-1}(A),
 \ee
where we defined the Chern-Simons form
 \be
 Q_{2n-1}(A)\equiv Q_{2n-1}(A,0) = n\int_0^1 dt\, P(A,F_t^{n-1}), \label{alan:chern-simons}
 \ee
and now,
 \be
 A_t = tA, \quad F_t = t\,dA + t^2 A\wedge A = tF + (t^2 - t)A\wedge A
 \ee
We see that, given an invariant polynomial $P_n(F)$, we can always
construct the associated Chern-Simons form $Q_{2n-1}(A)$ from
(\ref{alan:chern-simons}). Note that $P_n(F)$ is a $2n$-form, while
$Q_{2n-1}$ is a $(2n-1)$-form.

Since an invariant polynomial in $F$, $P_n(F)$, is closed and
generically non-trivial, it represents a non-trivial (de Rham)
cohomology class $[P_n(F)]\in H^{2n}(M,\mathbb{R})$, which is called a {\bf characteristic class}.
Since we have shown that the difference of two invariant polynomials
defined with respect to two different connections is exact, we have
by Stoke's theorem and for a manifold $M$ without boundary,
$\partial M = 0$,
 \be
 \int_M P_n(F')- \int_M P_n(F) = \int_M dQ_{2n-1}(A',A) = \int_{\partial M} Q_{2n-1}(A',A) = 0.
 \ee
This means that the integrals or periods $([P_n(F)],M)$ of these classes, usually called
characteristic numbers, do not depend on the connection chosen, in
other words, they are characteristic of the bundle itself
(transition functions)! This makes characteristic classes very
interesting objects to study the topology of fibre bundles. In
contrast, Chern-Simons forms obviously do depend on the connection
chosen, they are not even gauge invariant, but will prove to be very
useful nonetheless. We now go on to define some examples of
characteristic classes and Chern-Simons forms which are useful in
the study of gauge theories.

\subsection{Chern classes and Chern characters}

Let $P$ be a principal bundle, with structure group
$G=GL(k,\mathbb{C})$ or a subgroup thereof ($U(k)$,
SU(k),...)\footnote{One can equally well take the bundle to be an
associated complex vector bundle $E$ with one of the mentioned
structure groups.}. The {\bf total Chern class} is defined by (the
normalization of $F$ is for later convenience)
 \be
 c(F)=\det\left( 1 + \frac{i}{2\pi}F \right).
 \ee
Since $F$ is a 2-form, $c(F)$ is a sum of forms of even degrees,
 \be
 c(F) = 1 + c_1(F) + c_2(F)+...
 \ee
where $c_n(F)\in \Lambda^{2n}M$ is called the $n$-th {\bf Chern
class}\footnote{Strictly speaking, this is a representative of the
$n$-th Chern class, but we will follow the rest of the world in
calling these Chern classes by themselves.}. If dim$M=m$, all Chern
classes, $c_n(F)$ of degree $2n>m$, vanish. In general it can be
quite cumbersome to compute this determinant for higher dimensional
manifolds. Therefore we will diagonalize the matrix $\frac{i}{2\pi}
F$ (if for instance $G=SU(k)$, $F$ is anti-hermitian, so $iF$ is
hermitian and can be diagonalized by an $SU(k)$ rotation $g$) to a
matrix $\widetilde{F}$, with 2-forms $x_i$ on the diagonal. This
leads to
 \be
 \det (1+\widetilde{F}) &=& \det[\mbox{diag}(1+x_1,...,1+x_k)] = \prod_{i=1}^k(1+x_i) \nonumber\\
 &=& 1 + (x_1 + ... + x_k) + (x_1x_2 + ... + x_{k-1}x_k) + ... + (x_1x_2...x_k) \nonumber\\
 &=& 1 + \Tr \widetilde{F} + \frac 1 2 \left[ (\Tr \widetilde{F})^2 -\Tr \widetilde{F}^2\right]+...+\det \widetilde{F}.
 \ee
Note that in the second line we encounter the elementary symmetric
functions of $\{x_i\}$ and that all manipulations are well defined
since the $x_i$ are 2-forms and thus commute (the wedge product is
always understood). For an invariant polynomial
$P_n(F)=P_n(g^{-1}Fg)=P_n(2\pi \widetilde{F}/i)$ (note that the
trace always guaranties invariance), so we find the following
expressions for the Chern classes:
 \be
 c_1(F) &=& \Tr \widetilde{F} = \frac{i}{2\pi} \Tr F \\
 c_2(F) &=& \frac 1 2 \left[ (\Tr \widetilde{F})^2 -\Tr \widetilde{F}^2\right] = \frac{1}{8\pi^2}
 \left[ \Tr(F\wedge F)- \Tr F \wedge \Tr F \right] \\
 &\vdots& \nonumber\\
 c_k(F) &=& \det \widetilde{F} = \left(\frac{i}{2\pi}\right)^k \det F
 \ee
To show how the computation of Chern-Simons forms goes about, let's
start with a ridiculously easy example. Consider a $U(1)$-bundle
over some 2-dimensional manifold. The only Chern class which can be
defined is $c_1(F)$ and obviously, since locally $F=dA$, we find
 \be
 c_1(F) = d\left( \frac{i}{2\pi} A \right),\quad \mbox{so that}\quad Q_1(A) = \frac{i}{2\pi} A.
 \ee
Killing a fly with a jackhammer, we now use formula
(\ref{alan:chern-simons}) to compute the same thing
 \be
 Q_1(A) = \int_0^1 dt\, P(A) = \int_0^1 dt\, c_1(A) = \int_0^1 dt\, \frac{i}{2\pi} A = \frac{i}{2\pi} A.
 \ee

Now that we have earned some trust in (\ref{alan:chern-simons}), we
compute something less trivial. Consider an $SU(2)$-bundle over a
4-dimensional manifold. Since for $SU(k)$ we have that $\Tr F = 0$,
the first Chern class vanishes. Let's try to compute the
Chern-Simons form related to the second Chern class,
 \be
 Q_3(A) &=& 2 \int_0^1 dt\, P(A,F_t) = \frac{1}{4\pi^2} \int_0^1 dt\, \Tr (A\wedge F_t) \nonumber\\
 &=& \frac{1}{4\pi^2} \int_0^1 dt\, \Tr (t A\wedge dA + t^2 A\wedge A\wedge A) \nonumber\\
 &=& \frac{1}{8\pi^2} \Tr \left(A\wedge dA + \frac 2 3 A\wedge A\wedge A\right).
 \ee
which is of course the most famous example of a Chern-Simons form in
physics.

Since the periods of Chern classes are independent of the
connection, these numbers, called {\bf Chern numbers}, are denoted
as
 \be
 c_n \equiv ([c_n(F)],M) = \int_M c_n(F).
 \ee
One can show that on a compact manifold, these numbers are always
integers, $c_n = k$, a phenomenon called topological quantization.
We will see instances of this where the Chern numbers compute the
monopole charge or instanton number later.

We will now briefly discuss another characteristic class called the
Chern character, because it has some properties which make it easier
to compute than Chern classes (one can afterwards compute the Chern
classes from the Chern characters) and because it appears in the
Atiyah-Singer index theorem. The {\bf total Chern character} (again
for $G \subseteq GL(k,\mathbb{C})$) is defined by
 \be
 ch(F) = \Tr \exp \left( \frac{i}{2\pi}F \right) = \sum_{n} \frac{1}{n!}\Tr \left( \frac{i}{2\pi}F \right)^n.
 \ee
This is again a sum over even forms, the {\bf Chern characters}
 \be
 ch_n(F) = \frac{1}{n!}\Tr \left( \frac{i}{2\pi}F \right)^n
 \ee
By again diagonalizing $iF$ to a diagonal matrix $\widetilde{F}$
with eigenvalues $\{x_i\}$, using
 \be
 \Tr \exp(\widetilde{F}) = \sum_{i=1}^k \exp(x_i) = \sum_{i=1}^k (1 + x_i + \frac 1 2 x_i^2+...),
 \ee
and expressing the result in terms of elementary symmetric functions
of $\{x_i\}$, one can relate the Chern characters to the Chern
classes. A few examples are,
 \be
 ch_0(F) &=& k \\
 ch_1(F) &=& c_1(F) \\
 ch_2(F) &=& -c_2(F) + \frac 1 2 c_1(F)\wedge c_1(F).
 \ee
Since for a Dirac monopole we only need $c_1(F) = ch_1(F)$ and for
$SU(2)$-instantons $c_1(F)=0$, so that $ch_2(F)=-c_2(F)$, we will
not really see the difference between the two. We will mostly refer
to the Chern class, when speaking about either of the two.

\section{Some applications}

We will now apply the formalism we developed to two standard
examples of the topology of gauge bundles, Dirac monopoles and
instantons. We will not talk about (the more interesting)
non-Abelian 't Hooft-Polyakov monopoles, because these do not appear
in pure gauge theory, but require a Higgs field, which is a section
of an associated vector bundle and are not as such `pure' examples
of the topology of gauge bundles. For more on these and other
applications of bundles to physics, see \cite{Nakahara} -
\cite{Nash}.

\subsection{Dirac monopoles}
\label{alan:monopole}

Consider a magnetic monopole in Maxwell theory (Abelian) at the
origin of 3-dimensional Euclidean space, $\mathbb{R}^3$. If $q$ is
the magnetic charge of the monopole, we can take the magnetic charge
distribution to be $\rho(x) = 4\pi q \delta(x)$. Let $B^i$ be the
components of the magnetic field (a vector field on $\mathbb{R}^3$).
From Maxwell's equations, we know that $\partial_i B_i(x) = q
\delta(x)$, which has the spherically symmetric solution
 \be
 B = \frac{q}{r^2}\frac{\partial}{\partial r}.
 \label{alan:vectormonopole}
 \ee
This expression becomes infinite at the origin, so that strictly
speaking it is only a physically relevant solution on
$\mathbb{R}^3_0 = \mathbb{R}^3\backslash \{(0,0,0)\}$. There are no
non-trivial bundles over $\mathbb{R}^3$, but for $\mathbb{R}^3_0$
there is the possibility that there exists a fibre bundle
description of this kind of monopole. Since $\mathbb{R}^3_0$ is
homotopic to a sphere and we are considering a pure $U(1)$-gauge
theory, the proper bundle setting is that of a principal
$U(1)$-bundle over $S^2$, $P(U(1),S^2)$.

We described two ways for classifying this bundle. One was the
characterization of $\Pi_1(U(1))$ by the topological degree or
winding number of the map from the equator $S^1$ of $S^2$ to the
fibre $U(1)$. The other was by computing the first Chern class
$c_1(F)$ of a $U(1)$-connection on the sphere $S^2$.

Let's look at the second approach. The sphere can be covered by two
patches $U_N$ and $U_S$, corresponding to the northern and southern
hemisphere respectively, with $U_N \cap U_S = S^1$. On each patch we
have a local 1-form gauge potential, $A^N$ on $U_N$ and $A^S$ on
$U_S$. Since we are dealing with an Abelian structure group, the
2-form field strength $F$ is gauge invariant. This means that on the
equator
 \be
 F\vert_{S^1} = dA^N = dA^S.
 \ee
The first Chern number is computed as follows:
 \be
 c_1 = \int_{S^2}c_1(F) = \frac{i}{2\pi} \int_{S^2} F = \frac{i}{2\pi}\left( \int_{U_N}
 F + \int_{U_S} F \right).
 \ee
The field strength is locally exact on both hemispheres, so by
Stoke's theorem we find,
 \be
 c_1 = \frac{i}{2\pi}\left( \int_{\partial U_N} A^N + \int_{\partial U_S} A^S
 \right) = \frac{i}{2\pi}\int_{S^1}\left( A^N - A^S \right).
 \ee
On the equator both potential 1-forms are related by a transition
function $g\in U(1)$,
 \be
 A^S = A^N + g^{-1}dg.\label{alan:fromNtoS}
 \ee
This leads to
 \be
 c_1 = \frac{1}{2\pi i} \int_{S^1}g^{-1}dg,
 \ee
which we recognize as the winding number of the map $g:S^1
\rightarrow U(1)$. We see that for a map $g_n$ of winding number
$n$, we find that $c_1 = \mbox{deg}(g_n) = n$.

To connect to physics, we now write down an explicit solution. The
1-form $\beta$ associated to the vector field $B$ in
(\ref{alan:vectormonopole}) is (we use spherical coordinates
 with metric components $\eta_{rr}=1$, $\eta_{\theta\theta}=r$ and $\eta_{\varphi\varphi}=r\sin\theta$),
 \be
 \beta = \frac{q}{r^2}dr.
 \ee
The field strength 2-form $F$ is the Hodge dual to this (to find
complete agreement with the general theory, we need to include the
Lie algebra factor $i$),
 \be
 F = i*\beta = i\sqrt{\det\eta}\,B_r\, \varepsilon^{r}_{~\theta \varphi}\,d\theta \wedge
 d\varphi = iq \sin\theta\, d\theta \wedge d\varphi.
 \ee
This is the field strength 2-form which represents a Dirac monopole
of charge $q$. A possible gauge potential 1-form that leads to this
field strength is $A = -i\cos\theta\, d\varphi$. Since spherical
coordinates are badly behaved along the entire $z$-axis ($\theta =
0$, $\pi$), we can't use this potential for either hemisphere. We
can however define,
 \be
 A^N &=& iq(1-\cos\theta)d\varphi \quad \mbox{on} \quad U_N\\
 A^S &=& -iq(1+\cos\theta)d\varphi \quad \mbox{on} \quad U_S,
 \ee
which are well defined on their respective patches and lead to the
same $F$. On the equator, we find
 \be
 A^N = A^S + 2iqd\varphi.
 \ee
This completely agrees with equation (\ref{alan:fromNtoS}) for
 \be
 g_n = e^{in\varphi}, \quad n = 2q,
 \ee
so that a monopole of charge $q$ corresponds to a $U(1)$-bundle over
$S^2$ with winding number $n = 2q$, or, to put it differently,
corresponds to an element $[2q]$ of $\Pi_1(U(1))$.

We conclude this subsection with an important note. The above
reasoning might falsely cause one to believe that one does not need
quantum mechanics to prove the quantization of magnetic charge. The
main assumption we used, however, is that one can use a bundle to
describe the magnetic field on a sphere in the first place. This
manifests itself in a gauge transformation by $g^{-1}dg$ in eq.
(\ref{alan:fromNtoS}), instead of by just a general closed one-form
on the equator. This is equivalent to the assumption that the
magnetic field is described by an integral 2-form (a first Chern
class) as opposed to a generic 2-form, which is not integral. The
integrality of the magnetic field of a monopole is usually proved by
considering the wave function of a quantum mechanical particle in
the neighborhood of the monopole. To summarize: classically, the
magnetic field is just a (non-integral) 2-form (which consequently
is not the curvature of a bundle), while quantum mechanically, it is
an integral 2-form, which means that it can be seen as the curvature
(first Chern class) of a bundle.

\subsection{Holonomy and the Aharonov-Bohm effect}

We already briefly discussed holonomy in subsection
\ref{alan:parallel}, but let us come back to it in little more
detail. Consider a principal bundle $P(M,G)$ with a connection
1-form $\omega$. Let $\gamma$ be a curve on $M$ and $\gamma_P$ be a
horizontal lift. Suppose for the moment that $\gamma$ is contained
within a single patch $U$ and let $s: U \rightarrow P$ be a section
on $U$. This means that $\gamma_P(t) = s(\gamma(t)) g(t)$. The aim
is to compute $g(t)$ to have a local description of parallel
transport (local because this description depends on $s$).

If $X\in T\gamma M$ is the tangent to $\gamma$, the tangent to
$\gamma_P$ is $X_P = \gamma_{P*}X\in T_{\gamma_P}P$. Since
$\gamma_P$ is horizonal, we have that $X_P\in H_{\gamma_P}P$ or
$\omega(X_P)=0$. According to (\ref{alan:from_A_to_w}), this means
 \be
 g^{-1}\pi^* A(X_P)g + g^{-1}d_P g(X_P) = 0.
 \ee
Using that $\pi_* X_P = \pi_* \gamma_{P*}X = X$, we find on $M$,
 \be
 g^{-1}A(X)g + g^{-1}d g(X) = 0,
 \ee
or
 \be
 d g(X) = X(g) = -A(X)g.
 \ee
Since $X$ is tangent to $\gamma$, we have
 \be
 X(g) = \frac{d}{dt}g(\gamma(t)) \quad
 \mbox{and} \quad A(X)= A_a X^a = A_a \frac{d}{dt}x^a(\gamma(t)),
 \ee
where $A = A_a dx^a$. Writing $g(\gamma(t)) = g(t)$ and
$x^a(\gamma(t))=x^a(t)$, this leads to
 \be
 \frac{d g(t)}{dt}g(t)^{-1} = -A_a \frac{d x^a(t)}{dt}.
 \ee
For $G=U(1)$ this has the solution (suppose that $g(0)=e$)
 \be
 g(t) = \exp \left[-\int_0^t dt\, A_a \frac{d x^a(t)}{dt}\right] =
 \exp \left[-\int_{\gamma(0)}^{\gamma(t)} A_a d x^a\right]
 \ee
For short, we can write
 \be
 g(t) = \exp \left[-\int_{\gamma} A\right].
 \ee
For a non-Abelian structure (gauge) group, this is modified to
 \be
 g(t) = P \exp \left[-\int_{\gamma} A\right].
 \ee
where $P$ indicates that the exponential is defined by its power
series expansion and that the matrix-valued forms should always be
path ordered. In physics this is called a Wilson line.

If $s' = s h$ is another section on $U$ (or on an overlap with
another patch $U'$), related to $s$ by a group element $h\in G$, one
can show that if $\gamma_P(t) = s'(\gamma(t))g'(t)$, we find
 \be
 g'(t) = h^{-1}(t) g(t)h(0).
 \ee
This shows that if $G=U(1)$ and if $\gamma(0)=\gamma(1)$, so that
$h(1)\equiv h(\gamma(1))= h(\gamma(0)) = h(0)$, then
 \be
 g_\gamma \equiv \exp \left[-\oint_{\gamma} A\right],
 \ee
called a Wilson loop, is gauge invariant. $g_\gamma$ is nothing but
an element of the holonomy group $Hol(P)$ we discussed in subsection
\ref{alan:parallel}. We see that in the non-Abelian case this
procedure doesn't lead to a gauge invariant quantity, but
 \be
 g_\gamma' = h^{-1}g_\gamma h, \quad h = h(0) = h(1).
 \ee
If we take the trace of this though, we do get a gauge invariant
quantity. In non-Abelian gauge theories the Wilson loop is thus
defined as the trace of the Wilson line around a closed loop,
 \be
 W_\gamma = \Tr g_\gamma = \Tr P \exp \left[-\oint_{\gamma} A\right],
 \ee
and equals the trace of the holonomy at $p=\gamma_P(0)$.

To illustrate this, consider a solenoid along the $x_3$-axis in
$\mathbb{R}^3$. The $U(1)$ magnetic field is uniform in the interior
of the solenoid and practically vanishing outside of it. In the
limit of infinitely thin and long solenoid, the magnetic field is
strictly $0$ in the exterior region, which is $\mathbb{R}^3_0$, but
there still is a non-zero flux $\Phi$ associated to it. By Stoke's
theorem, this means that the gauge potential $A$ cannot be zero in
the exterior region, since for any curve $\gamma$ encircling the
solenoid, which spans a surface $A_\gamma$
 \be
 \oint_\gamma A = \int_{A_\gamma}dA = \int_{A_\gamma}F = \Phi.
 \ee
Since $F=0$ outside of the solenoid, there is no classical effect on
a particle moving alongside it. Quantum mechanically however, it is
known that $A$ can have a physical meaning. To see this, consider a
wave function $\psi$ of a particle moving in the $x_1x_2$-plane
perpendicular to the solenoid. This is described by a section of a
complex line bundle over $\mathbb{R}^2_0$ associated to the
principal bundle $P(\mathbb{R}^2_0, U(1))$ by the obvious
representation
 \be
 \psi \rightarrow \rho(e^{\alpha})\psi = e^{\alpha}\psi.
 \ee
In a path integral approach, every path $\gamma$ is weighed by a
factor
 \be
 g(t) = e^{iS_\gamma}, \quad S_\gamma = \int_\gamma dt\, L,
 \ee
where the important part of the Lagrangian $L$ is the part involving
the gauge potential,
 \be
 L = A_i(x) \frac{d x^i}{dt}.
 \ee
In other words, every path is weighed by a factor (note that we
absorbed the Lie algebra factor into $A$ to make contact with our
formalism),
 \be
 g(t) = \exp \int_\gamma A.
 \ee
In a double slit experiment, with the solenoid placed between the
slits, part of the wave function $\psi_1$ will move along path
$\gamma_1$ above the solenoid and part $\psi_2$ will move along a
path $\gamma_2$ underneath it. The total wave function
is\footnote{We are mixing path integral and ordinary QM arguments.
Since the only real importance is wether the path passes underneath
or above the solenoid, the path integral essentially reduces to the
sum of 2 paths, so that both viewpoints essentially give the same
information.}
 \be
 \psi &=& \exp \left[\int_{\gamma_1} A\right]\psi_1 + \exp \left[\int_{\gamma_2} A\right]
 \psi_2\nonumber\\
 &=& \exp \left[\int_{\gamma_1} A\right]\left\{ \psi_1 + \exp \left[\int_{\gamma_2} A- \int_{\gamma_1} A\right]
 \psi_2\right\}.
 \ee
What is important of course, is the phase difference,
 \be
 \int_{\gamma_2} A- \int_{\gamma_1} A = \oint_{\gamma}A, \quad
 \gamma = \gamma_2 - \gamma_1.
 \ee
We see that the probability to find a particle at a certain point on
the screen is influenced by the gauge potential through the holonomy
of the connection
 \be
 \vert \psi \vert^2 = \vert \psi_1 + g_\gamma \psi_2 \vert^2 = \vert \psi_1 + e^\Phi \psi_2
 \vert^2.
 \ee
Note that the only reason why $\oint A$ can have physical meaning is
because it is gauge invariant (independent of local
trivializations).

$G$-bundles over a circle are classified by $\Pi_0(G)$, implying
that a $U(1)$-bundle over $S^1$ is necessarily trivial. Hence, this
example shows that a trivial bundle can contain non-trivial physics,
i.e. the Aharonov-Bohm effect.

\subsection{Instantons}
\label{alan:instanton}

Instantons are traditionally defined as smooth finite action
solutions of Yang-Mills theory on 4-dimensional Euclidian space
$\mathbb{R}^4$. We will only consider the case of $SU(2)$. There
exist no non-trivial bundles over $\mathbb{R}^4$, but the finiteness
of the action imposes boundary conditions at infinity, which allow
for the existence of topologically non-trivial solutions of the
field equations. This is seen as follows: To get a finite action,
the field strength has to go to zero (fast enough) at infinity. This
means that along a sufficiently large 3-sphere, $S^3_\infty$, the
gauge potential has to be pure gauge
 \be
 A\vert_{S^3_\infty} = g^{-1}dg\, \Longrightarrow\, F\vert_{S^3_\infty}
 = 0,
 \ee
so these solutions are classified by maps $g:S^3_\infty \rightarrow
SU(2)$. The reason for their stability is the fact that one cannot
change the homotopy class of this map while keeping the total action
finite. Computing the second Chern number, using that $c_2(F) =
dQ_3(A)$ (and assuming that there is no contribution outside of
$S^3_\infty$),
 \be
 c_2 &=& \int_{\mathbb{R}^4} c_2(F) = \int_{S^3_\infty} Q_{3}(A) =
 \frac{1}{8\pi^2}\int_{S^3_\infty} \Tr \left(A\wedge dA + \frac 2 3 A\wedge A\wedge
 A\right)\nonumber\\
 &=& \frac{1}{8\pi^2}\int_{S^3_\infty} \Tr \left(F\wedge A - \frac 1 3 A\wedge A\wedge
 A\right) = -\frac{1}{24\pi^2}\int_{S^3_\infty} \Tr \left( A\wedge A\wedge
 A\right)\nonumber\\
 &=& -\frac{1}{24\pi^2}\int_{S^3_\infty} \Tr \left( g^{-1}dg\wedge g^{-1}dg\wedge
 g^{-1}dg\right).
 \ee
Which is (up to a sign) exactly the topological degree of the map
$g:S^3_\infty \rightarrow SU(2)$.

To gain more control over the situation and allow for a bundle
description of instantons, we consider a one-point compactification
of $\mathbb{R}^4$ to $S^4$, by adding to it the point at infinity,
$\mathbb{R}^4 \cup \{\infty\}= S^4$. This means that we want to look
at principal $SU(2)$-bundles over $S^4$, $P(SU(2),S^4)$. As we saw
in subsection \ref{alan:homotopy}, this is classified by
$\Pi_3(SU(2))$, while on the other hand one can compute the second
Chern number of an $SU(2)$-connection on $S^4$.

Again, we can cover the 4-sphere by two open sets, the northern and
southern hemisphere, $U_N$ and $U_S$ respectively. This time, the
field strength 2-form is not invariant, but
 \be
 F^N = dA^N + A^N\wedge A^N,\quad F^S = dA^S + A^S\wedge A^S,
 \ee
with
 \be
 A^N = g^{-1}A^Sg + g^{-1}dg \,\Rightarrow\, F^N = g^{-1}F^Sg.
 \label{alan:fromNtoS_nonab}
 \ee
Let's try to compute the second Chern number
 \be
 c_2 &=& \int_{S^4} c_2(F) = \int_{U_N}dQ_3(A^N) + \int_{U_S}dQ_3(A^S)
 = \int_{S^3} \left(Q_3(A^N) - Q_3(A^S)\right)\nonumber\\
 &=& \frac{1}{8\pi^2}\int_{S^3} \Tr\left(F^N\wedge A^N - \frac 1 3 A^N\wedge A^N\wedge A^N -
 F^S\wedge A^S + \frac 1 3 A^S\wedge A^S\wedge A^S\right)\nonumber\\
 &=& -\frac{1}{8\pi^2}\int_{S^3} \Tr\left(\frac 1 3 g^{-1}dg\wedge g^{-1}dg\wedge
 g^{-1}dg - d(g^{-1}A^N\wedge dg)\right)
 \ee
where in the last step we used the gauge transformations
(\ref{alan:fromNtoS_nonab}), the identity
 \be
 dg^{-1} = -g^{-1}dg g^{-1},
 \ee
and the fact that for three $\mathfrak{g}$-valued 1-forms
 \be
 \Tr(\alpha\wedge\beta\wedge\gamma) =
 \Tr(\gamma\wedge\alpha\wedge\beta).
 \ee
Since $S^3$ has no boundary, we arrive at the expected result
 \be
 c_2 = \frac{1}{24\pi^2}\int_{S^3} \Tr\left(g^{-1}dg\wedge g^{-1}dg\wedge
 g^{-1}dg \right).
 \ee
Again we find that $c_2 = \mbox{deg}(g)=k$, as described in
(\ref{alan:deg3sphere}). We conclude that the classification by the
second Chern class $[c_2(F)]$ is equivalent to the one given by
$\Pi_3(SU(2))$. Both are characterized by the degree of the
transition function of the bundle.

What does this have to do with instantons? As we already stated, we
have to find finite action solutions of the action for $F =
F_{ij}dx^i\wedge dx^j$, $F_{ij}=F_{ij}^{~a}\, T_a$
 \be
 S_E = \frac 1 4 \int dx^4\, F_{ij}^{~a} F^{ij}_{~a} =
 -\frac 1 2 \int dx^4\, \Tr \left(F_{ij} F^{ij}\right) = - \int \Tr (F\wedge
 *F),\label{alan:YMaction}
 \ee
where $*F$ is the Hodge dual of $F$, in flat space
 \be
 *F_{ij} = \frac 1 2 \varepsilon_{ijkl}F^{kl}.
 \ee
For $\mathfrak{su}(2)$, the Lie algebra of $SU(2)$, we take the
generators (in the defining representation) to be,
 \be
 T_a = \frac{1}{2i}\,\tau_a,
 \ee
where $\tau_a$, $a\in \{1,2,3\}$, are the Pauli matrices. The generators have the following properties
 \be
 \Tr(\tau_a \tau_b) = 2\delta_{ab}\, &\Rightarrow&\, \Tr(T_a T_b) =
 -\frac 1 2 \,\delta_{ab},\\
 {[\tau_a,\tau_b]} = 2i\, \varepsilon_{ab}^{~~\,c}\,\tau_c &\Rightarrow&
 [T_a,T_b] = \varepsilon_{ab}^{~~\,c}\,T_c.
 \ee

To write the action in a convenient way, we calculate the following positive definite
object
 \be
 \frac 1 4 \int(F_{ij}^{~a}\pm *F_{ij}^{~a})(F^{ij}_{~a}\pm *F^{ij}_{~a}) &=&
 -\int \Tr (F\pm*F)\wedge *(F\pm*F)\nonumber\\
 &=& -2 \int \Tr (F\wedge *F)\mp 2 \int \Tr (F\wedge F).
 \ee
From this, we find
 \be
 S_E &=& -\frac 1 2 \int \Tr (F\pm*F)\wedge *(F\pm*F) \pm \int \Tr (F\wedge
 F)\nonumber\\
 &\geq& 8\pi^2 \vert k\vert.\label{alan:lowerbound}
 \ee
Where we defined the {\bf instanton number} $k$ to be
 \be
 k=-c_2=ch_2.
 \ee
We see that the positive definite action $S_E$ is bounded from below and that a minimum is
attained when
 \be
 F = \pm *F,
 \ee
called self-dual (SD) and anti-self-dual (ASD) instantons
respectively. We have chosen the instanton number $k$ in such a way
that it is positive (negative) for (A)SD instantons, as one can see
from (\ref{alan:YMaction}). This shows that (\ref{alan:lowerbound})
does establish a lower bound. Since (A)SD instantons are minima of
the action, they are solutions to the equations of motion.

\subsection{Further applications and remarks}

Characteristic classes are a very important ingredient for the
Atiyah-Singer index theorem. It would take us too far to go into a
lot of details, but we will try to sketch some aspects of this
application of characteristic classes to the study of anomalies and
moduli spaces. For more information on the use of fibre bundles in
the study of anomalies, see \cite{Nakahara} and \cite{Bertlmann}

The idea is to compute an analytic index of certain differential
operators defined on a bundle by computing topological quantities of
the bundle expressed as integrals of characteristic classes of the
bundle. By an analytic index of an operator $D$, we mean
 \be
 \mbox{Ind} D = \mbox{dim ker} D - \mbox{dim ker} D^{\dag}
 \ee
where $D^{\dag}$ is the adjoint of $D$ with respect to an inner
product,
 \be
 (u,Dv) = (D^{\dag}u,v).
 \ee
For the analytic index to be well defined, we need that both ker $D$
and ker $D^{\dag}$ are finite dimensional. A (bounded) operator
which satisfies these conditions is called a Fredholm operator. One
can show that certain differential operators on compact boundaryless
manifolds, called elliptic operators, are always Fredholm operators.

It is for these elliptic differential operators that Atiyah and
Singer found another way to express the index, namely in terms of
characteristic classes. One important example of where this is
relevant in physics, is for the chiral anomaly. Consider a massless
Dirac spinor coupled to an $SU(2)$ gauge theory on a 4-dimensional
manifold $M$. Classically, there is a global chiral symmetry
 \be
 \psi' = e^{i\gamma_5 \alpha}\psi, \quad \bar{\psi}' = \bar{\psi} e^{i\gamma_5
 \alpha},
 \ee
which leads to a conserved current
 \be
 \partial_a j_5^a = 0.\label{alan:conserved_chiral}
 \ee
Quantum mechanically this symmetry is broken, so that it is
anomalous. One can show that the right hand side of
(\ref{alan:conserved_chiral}) is no longer zero, but that its
integral over $M$ is given by the index of the Dirac operator
 \be
 \mbox{ind}(i\displaystyle{\not}\nabla_+) = \mbox{dim ker}
 (i\displaystyle{\not}\nabla_+) - \mbox{dim ker}
 (i\displaystyle{\not}\nabla_-) = n_+ - n_-\,
 \ee
where
 \be
 i\displaystyle{\not}\nabla_\pm = i\gamma^a \nabla_a P_+ = i \gamma^a (\partial_a +
 A_a)\frac 1 2 (1 \pm \gamma_5),
 \ee
and $n_\pm$ are the number of positive and negative chirality zero
modes of the Dirac operator, respectively. It is clear that if the
index of this operator is nonzero, the chiral symmetry is broken.
For this example, the Atiyah-Singer index theorem states that (if
all relevant characteristic classes of the tangent bundle $TM$ are
zero),
 \be
 \mbox{ind}(i\displaystyle{\not}\nabla_+) = \int_M ch_2(F) = ch_2,
 \ee
so the index is given by the second Chern character of $P$. Since
for $SU(2)$ this is equal to the instanton number $k$, the statement
becomes,
 \be
 \int_M dx^4\, \partial_a j_5^a =
 \mbox{ind}(i\displaystyle{\not}\nabla_+) = n_+ - n_- = ch_2 =
 k.\label{alan:chiral_anomaly}
 \ee
We see that the index computes the anomaly of the theory and shows
the obstruction for the classical symmetry to become a quantum
symmetry. Moreover, (\ref{alan:chiral_anomaly}) shows that the
instanton background breaks the symmetry. On the other hand if there
are no instantons, the chiral symmetry is a symmetry in the full
quantum theory. Because the current in this case carries no group
index, this is also called the Abelian anomaly. A similar discussion
can be given for a non-Abelian anomaly.

The index theorem can for instance also be used to compute the
dimension of the moduli space of instantons in $SU(N)$. One can show
that the number of parameters to describe a general instanton (for a
given winding number $k$) is related to the number of zero modes of
a kind of Dirac operator. It would take us too far to discuss this
in detail, but an index slightly different from the above one
computes the number of zero modes of this operator. Using this, one
can show that the moduli space of $SU(N)$ instantons with winding
number $k$ is $4kN$-dimensional \cite{Instantons}.

We have seen two examples of integral cohomology classes at work,
$[c_1(F)]$ for the monopole and $[c_2(F)]$ for the instanton. Both
are elements of $H^p(M,\mathbb{Z})$, where dim$M = p$ ($p=2$ for the
monopole and $p=4$ for the instanton). They both represent an
obstruction for the principal bundle $P$ to be trivial. In this
sense it's clear that they both represent an obstruction for $P$ to
have a section, this can only happen if $P$ is trivial. Note also
that in both cases, $\mathbb{Z}$ turns out to be $\Pi_{p-1}(G)$,
namely $\Pi_1(U(1)) = \Pi_3(SU(2)) = \mathbb{Z}$. These turn out to
be special cases of a far more general result.

Let $P(M,G)$ be a principal bundle, and let $M_p$ be the
$p$-dimensional skeleton of $M$. Define $s_p$ to be a section of the
bundle $P$ defined only over the skeleton, i.e. $s_p: M_p
\rightarrow \pi^{-1}(M_p)$. Generically, any given $s_p$ can be
extended to a section $s_{p+1}$ over $M_{p+1}$. However, if $P(M,G)$
is not trivial, there will be obstructions to continuing such a
chain of extensions. Without intending to be fully rigorous, the
general result found in \cite{Steenrod} can be stated as follows:
\begin{quote}
Let $p$ be the first dimension such that sections over $M_{p-1}$
cannot be extended to sections over $M_p$. Then, the obstruction to
building a section of $P$ over an $p$-dimensional skeleton of $M$
lies in the cohomology group $H^p(M, \Pi_{p-1}(G))$.
\end{quote}

We will not try to delve much deeper into these very interesting
matters, but note that $\Pi_{p-1}(G)$ need not be $\mathbb{Z}$. So
the relevant cohomology classes need not be integer valued, but
might be, for instance, $\mathbb{Z}_2$ valued, like Stiefel-Whitney
classes. For instance, the non-triviality of the M\"obius strip is
measured by the first Stiefel-Whitney class, which belongs to the
cohomology group \be H^1(S^1, \Pi_{0}(\mathbb{Z}_2)) = H^1(S^1,
\mathbb{Z}_2) = \mathbb{Z}_2. \ee This tells us that there are two
ways to make a $\mathbb{Z}_2$ bundle over $S^1$.

The two physics examples we studied can be interpreted as
follows: For the monopole, we had
 \be
 H^2(S^2, \Pi_{1}(U(1))) = H^2(S^2, \mathbb{Z}) = \mathbb{Z}.
 \ee
This means that, if the class is not the trivial element in
cohomology, there is an obstruction to finding a section over a
2-dimensional skeleton of $S^2$. In this case, because this is
incidentally the top class of the manifold, this means indeed that
one cannot construct a section of $P(S^2,U(1))$, so that the bundle
is non-trivial. A charge $n$ monopole corresponds to a connection on
a bundle which is in the class $[n]$ of $H^2(S^2, \Pi_{1}(U(1)))$,
where $n$ is the degree of the appropriate map or is the first Chern
number $c_1$ (see subsection \ref{alan:monopole}).

For instantons, the story is analogous. In this case the relevant
object is,
 \be
 H^4(S^4, \Pi_{3}(SU(2))) = H^4(S^4, \mathbb{Z}) = \mathbb{Z}.
 \ee
This again describes the possible obstruction to finding a section
of $P(S^4,SU(2))$, because this is again the top class. The
instanton with instanton number $k$ corresponds to a connection on a
bundle in class $[k]$, where $k = -c_2 = ch_2$ or is again the
degree of the appropriate map (see subsection \ref{alan:instanton}).
Note that for both the monopole and instanton, the gauge potential
describes one possible connection on the bundle. There are many
possible connection on a bundle in a certain class $[n]$, but they
will all compute the same Chern class and be characterized by the
same topological degree. In this sense, the bundle itself is more
fundamental than the specific solution of the gauge theory we
construct. Of course, this does not take away the meaning of these
specific solutions. For instance, an instanton is still a minimum of
the action, while other gauge configurations in the same topological
class are generically not.

Notice that a characteristic class can only be viewed as an element
of the group $H^p(M, \Pi_{p-1}(G))$ if it represents the
\emph{first} obstruction to the extendibility of sections over
skeletons. Higher obstructions require a different interpretation.
For example, if $H^2(M,\mathbb{Z})$ is not trivial for some
four-dimensional $M$, then it is possible to have a non-trivial
$U(1)$ bundle over $M$ with non-zero first Chern class $F$. Although
the second Chern class of a rank one bundle is always trivial, it is
possible in this case to have a non-trivial second Chern character.
This means that one can have a $U(1)$ instanton on a
four-dimensional manifold even though $\Pi_3(U(1))=0$. This is
possible because the second Chern character does not represent the
first obstruction of the $U(1)$ bundle, and hence does not lie in
$H^4(M, \Pi_3(U(1))$. Instead, in this case it lies in $H^4(M,
\Pi_1(U(1))$. Although such instantons are not wide-spread objects
in quantum field theory, they certainly make their appearance in
string theory. For those who are familiar with string theory, the
example in question is a D4 brane wrapped on a manifold with $H^2
\neq 0$, that carries lower-dimensional D2 and D0 charge. The D2
brane can be viewed as vortex or string living on the D4, whose
charge is given by the first Chern class of the $U(1)$ bundle on the
D4. The D0 brane can be viewed as a $U(1)$ instanton on the D4 (if
one ignores the time direction), such that its charge is given by
the second Chern character of the $U(1)$ bundle.

\bigskip

\noindent {\large {\bf Acknowledgments}}

\bigskip

\noindent We thank, of course, the organizers of the Second Modave
School on Mathematical Physics for providing us with the opportunity
to give this set of lectures and for creating the perfect
environment for doing so. We also thank the participants of the
Modave School for their useful questions and comments, in particular
Jarah Evslin and Laurent Claessens. Both authors are supported in
part by the Belgian Federal Science Policy Office through the
Interuniversity Attraction Pole P5/27 and in part by the European
Commission FP6 RTN programme MRTN-CT-2004-005104. AW is supported in
part by the ``FWO-Vlaanderen'' through project G.0428.06.

\renewcommand{\thesection}{\Alph{section}}
\setcounter{section}{0}

\section{Some differential geometry}

In this appendix we will quickly review some of the concepts of the
theory of differentiable manifolds that we will use in the rest of
these notes. We will also establish a lot of notation used throughout the main text.

\subsection{Manifolds and tangent spaces}
\label{alan:manifolds}

First of all, a differentiable manifold is roughly a smooth
topological space, which locally looks like $\mathbb{R}^n$. By
identifying an open subset of a manifold with an open subset of
$\mathbb{R}^n$, the notion of differentiability of a function from
$\mathbb{R}^n$ to $\mathbb{R}^m$ is passed on to one of a function
from one manifold to another. This means that one can compare
manifolds as smooth spaces. More importantly for these lectures, it
allows for doing physics on them, much like on $\mathbb{R}^n$. Let's
be a bit more concrete.

\begin{defi}
We call $M$ a differentiable manifold if the following conditions
are satisfied
\renewcommand{\labelenumi}{(\roman{enumi})}
\begin{enumerate}
\item $M$ is a topological space
\item $M$ is equipped with a set of pairs $\{(U_i,\varphi_i)\}$,
where $\{U_i\}$ is an open cover of $M$ (all $U_i$ are open sets and
$M=\bigcup_i U_i$) and $\varphi_i$ is a homeomorphism from $U_i$ to
an open subset of $\mathbb{R}^n$. $n$ is called the dimension of $M$
\item On an overlap $U_i \cap U_j \neq \emptyset$, the map $\varphi_j\circ
\varphi_i^{-1}:\mathbb{R}^n \rightarrow \mathbb{R}^n$ is
continuously differentiable.
\end{enumerate}
\end{defi}

Again, given a local chart $(U_i,\varphi_i)$ one can define physical
quantities on $U_i$ much like one would do on $\mathbb{R}^n$, where
the $\varphi_i$ define coordinates on $U_i$. The different patches
$U_i$ can however be glued together in a nontrivial way by the
transition functions $\varphi_j\,\varphi_i^{-1}$, so that globally a
manifold is a generalization of $\mathbb{R}^n$. We give this
definition for completeness, but to keep the notation tractable we
will be a bit sloppy throughout the text and keep the $\varphi_i$
implicit. For instance, to define the derivative of a function
$f:U_i\rightarrow \mathbb{R}$ at a point $x\in U_i$, one would have
to consider instead $f\circ \varphi_i^{-1}: \mathbb{R}^n \rightarrow
\mathbb{R}$ and use the definition of a derivative of a functional
on $\mathbb{R}^n$. We will omit $\varphi$ from this expression and
treat $f$ as though it were a function on $\mathbb{R}^n$.

First of all, we define the tangent space to $x\in M$.

\begin{defi}
Let $\gamma:[0,1]\rightarrow U_i:t\mapsto \gamma(t)$ be a curve on a
chart of $M$, such that $\gamma(0) = x$. A vector $X$ at $x$ tangent
to the curve $\gamma$ is called a tangent vector to $M$ at $x$. If
$\{x^a\}$ are a set of coordinates on $U_i$, $X$ can be represented
by the components
 \be
 X^a =\left. \frac{d}{dt}x^a(\gamma(t))\right\vert_{t=0}.
 \ee
The collection of the vectors at $x$ tangent to al curves that go
through $x$ is called the tangent space $T_xM$ at $x$.
\end{defi}

In a lot of practical situations (and to have an explicit
representation) it can be convenient to define a tangent vector by using a
function on $M$. Let $f$ be a function from $M$ to $\mathbb{R}$. One
defines a tangent vector by
 \be
 X (f) = \left. \frac{d}{dt} f(\gamma(t))\right\vert_{t=0},
 \label{alan:def2tangent}
 \ee
where now $X$ is represented by a differential operator
 \be
 X = X^a \frac{\partial}{\partial x^a}\equiv X^a \partial_a,
 \ee
so that the set $\{ \partial_a\}$ can be considered as a basis for
$T_xM$. Note that this definition is consistent with the first one,
since if we take the function $f$ to be the coordinate map $x^a$
that maps every point to its $a$-th coordinate, we get from
(\ref{alan:def2tangent}) that $X (x^a) = X^a$. A smooth assignment
of a tangent vector at every point $x\in M$ is called a vector field
$X(x)$ on $M$. In subsection \ref{alan:triviality} this is called a
section of TM, $X(x) \in \Gamma(M,TM)$.

\subsection{Differential forms}
\label{alan:forms}

\begin{defi}
The space dual to the tangent space $T_xM$ is the cotangent space
$T_x^*M$, that is, $T_x^*M$ is the space of all functionals from
$T_xM$ to $\mathbb{R}$. An element of $T_x^*M$ is called a cotangent
vector or a 1-form $\alpha$.
\end{defi}

The dual basis to $\{ \partial_a\}$ is denoted by $\{dx^a\}$, so
that we have that $dx^a(\partial_b)=\delta^a_b$. More generally,
this leads to
 \be
 \alpha(X) = \alpha_a dx^a (X^b \partial_b)=\alpha_a X^a.
 \label{alan:formonvector}
 \ee
In general a differential form of degree $p$ is an element of the
totally anti-symmetric tensor product of $p$ copies of $T_x^*M$.
This is accomplished by introducing the wedge product
 \be
 dx^a \wedge dx^b = - dx^b \wedge dx^a.
 \ee
Continuing this process, one can build a basis for the space of all
differential forms. The space of $p$-forms at $x$ is denoted by
$\Lambda_x^pM$. Again a smooth assignment of a 1-form at every point
of $M$, is called a 1-form $\alpha(x)$ on $M$ and is a section of
$T^*M$, $\alpha(x)\in \Gamma(M,T^*M) \equiv \Lambda M$. More
generally, a $p$-form on $M$ is an element of $\Lambda^p M$. A
general $\alpha_p\in \Lambda^pM$ can be expanded as
 \be
 \alpha_p = \frac{1}{p!}\, \alpha_{a_1...a_p}(x)\,dx^{a_1}\wedge ... \wedge
 dx^{a_p}. \label{alan:expansionform}
 \ee
In this way the product of two differential forms is defined, with
the property
 \be
 \alpha_p \wedge \beta_q = (-)^{pq} \beta_q \wedge \alpha_p,
 \ee
as can be seen from the expansion (\ref{alan:expansionform}). The
exterior differential is defined by
 \be
 d = \partial_a\, dx^a \wedge.
 \ee
which is a symbolic notation for
 \be
 d\alpha_p = \frac{1}{p!}\,\partial_a \alpha_{a_1...a_p} dx^a\wedge dx^{a_1}\wedge
 ... \wedge dx^{a_p}.
 \ee
This means that $d$ sends $p$-forms to $(p+1)$-forms, that it is
nilpotent, $d^2 = 0$, and that it is an anti-derivation
 \be
 d(\alpha_p\wedge\beta_q) = d\alpha_p \wedge \beta_q + (-)^p \alpha_p \wedge d\beta_q
 \ee
Note the very useful identity $X(f) = df(X)$, as is easily seen by
expanding both expressions. A similar expression for a 2-form
$\alpha$ is
 \be
 d\alpha(X,Y) = X(\alpha(Y)) - X(\alpha(Y)) -\alpha([X,Y])
 \ee
where $X$, $Y\in T_xM$ and $[X,Y] = [X,Y]^a \partial_a$ is the Lie
bracket,
 \be
 [X,Y](f)= X(Y(f)) - Y(X(f)).
 \ee
Differential forms are very important when it comes to defining
integration on a general manifold. On an $n$-dimensional manifold
$M$ an $n$-form $\alpha_n$ transforms as a volume element because of
the wedge product, so that the integral
 \be
 \int_M \alpha_n \equiv \int_M \alpha(x)\, dx^1 ...\; dx^n,
 \ee
is well defined. Here we used that a top form (of maximal dimension)
is always characterized by a single function,
 \be
 \alpha_n = \frac{1}{n!}\, \alpha_{a_1 ... a_n}(x) dx^{a_1} \wedge ...\wedge
 dx^{a_n} \Rightarrow \alpha_{a_1 ... a_n}(x) = \alpha(x)\varepsilon_{a_1 ...
 a_n},
 \ee
and
 \be
 \frac{1}{n!}\, \varepsilon_{a_1 ... a_n}dx^{a_1} \wedge ...\wedge dx^{a_n}
 = dx^1 \wedge ... \wedge dx^n
 \ee
Since we always work in Euclidean signature, the totally
ant-symmetric tensor is defined by $\varepsilon_{1...n}=1$ and
raising indices does not affect the sign. The most important tool we
will use, is Stoke's theorem,
 \be
 \int_M d\alpha = \int_{\partial M}\alpha,
 \ee
where $\partial M$ is the boundary of $M$ ($\partial\partial M =
0$).

A form $\alpha$ for which $d\alpha = 0$ is called closed, and if
$\alpha = d\beta$ for some form $\beta$, $\alpha$ is called exact. A
closed form is also called a cocycle and an exact form a coboundary.
It is clear that all exact forms are closed (because $d^2=0$), but
the reverse is not necessarily true on a non-trivial manifold. An
important theorem is Poincar\'e's lemma, which states that locally
(on an open set which is contractible) every closed form is exact.
The fact that on a general manifold this is not the case globally
means that the cohomology defined by closed forms that are not exact
contains a lot of information about the topology of a manifold. To
be more precise, one defines the group of $p$-cocycles
$Z^p(M,\mathbb{R})$ and the group of $p$-coboundaries
$B^p(M,\mathbb{R})$ as follows
 \be
 Z^p(M,\mathbb{R}) &=& \{ \alpha \in \Lambda^pM \vert d\alpha = 0 \},\\
 B^p(M,\mathbb{R}) &=& \{ \alpha \in \Lambda^pM \vert \alpha = d\beta,\, \beta\in \Lambda^{p-1}M \}.
 \ee
The $p$-th de Rham cohomology class is then defined as the quotient,
 \be
 H^p(M,\mathbb{R}) = Z^p(M,\mathbb{R}) /B^p(M,\mathbb{R}),
 \ee
so that a general element of $H^p(M,\mathbb{R})$ is an equivalence
class under the equivalence relation,
 \be
 \alpha_p \sim \beta_p \quad \mbox{iff} \quad \alpha_p = \beta_p + d\gamma_{p-1},
 \ee
and we denote their common equivalence class by $[\alpha_p] =
[\beta_p] \in H^p(M,\mathbb{R})$. A period of an element
$[\alpha]\in H^p(M,\mathbb{R})$ over a boundaryless submanifold $C
\in M$ is defined by
 \be
 ([\alpha], C) = \int_C \alpha,
 \ee
and because of Stoke's theorem this is independent of the choice of
representative of the equivalence class. Some cohomology classes,
like Chern classes (see section \ref{alan:topology}), are known to
have integral periods, $([c_n(F)],M) \in \mathbb{Z}$. We denote
these integral cohomology groups by $H^p(M,\mathbb{Z})$.

Note that if dim$M=m$, $\Lambda^pM$ and $\Lambda^{m-p}M$ have the
same dimension. Given a metric $g_{ab}$ on $M$, one can define an
isomorphism between the two called Hodge duality. When $\alpha_p$ is
given by (\ref{alan:expansionform}), its Hodge dual $*\alpha_{m-p}$
is defined by
 \be
 *\alpha_{m-p} = \frac{1}{p! (m-p)!}\,\sqrt{g}\, \alpha_{a_1...a_p}
 \varepsilon^{a_1...a_p}_{~~~~~~~a_{p+1}...a_m}\,dx^{a_{p+1}}\wedge ... \wedge dx^{a_m}.
 \ee
For Euclidean signature spaces, one finds $** = (-)^{p(m-p)}$. For
example, for a 2-form in 4-dimensional Euclidean space, we have
$**=1$. This means that a self-duality condition is well defined.

\subsection{Push-forwards and pull-backs}
\label{alan:pushandpull}

\begin{defi}
Given a map $f:M\rightarrow N$, there is always an induced map $f_*:
T_xM \rightarrow T_{f(x)}N$ called the push-forward (or differential
map) of $f$. This sends a vector $X$ tangent to a curve $\gamma$ at
$x=\gamma(0)\in M$ to a vector $f_*X$ tangent to the curve $f \circ
\gamma$ at the point $y=f(x)=f(\gamma(0))\in N$.
\end{defi}

Concretely, this means that, if $\{y^a\}$ are a set of coordinates
for $y\in V\subset N$, where $V$ contains $f(x)$ the components of
$f_*X$ are
 \be
 (f_*X)^a = \left. \frac{d}{dt}y^a (f(\gamma(t)))\right\vert_{t=0}.
 \ee
Again, sometimes it is convenient to define $f_*X$ by using an
auxiliary function $g:N\rightarrow \mathbb{R}$. A definition
equivalent to the previous one is
 \be
 f_*X(g) = X(g \circ f).
 \ee
From this definition it is easy to express $f_*X$ in terms of $X$.
First of all, using coordinate bases for the coordinates $\{x^a\}$
on $U\subset M$, where $U$ contains $x$, and the $\{y^a\}$ defined
previously, we find
 \be
 (f_*X)^b \frac{\partial}{\partial y^b}g(y) = X^b \frac{\partial}{\partial
 x^b}(g(f(x))).
 \ee
Note that on the left hand side, $g(y)$ means that $g$ is expressed
in the coordinates $\{y^a\}$, while on the right hand side,
$g(f(x))$ means that $g$ is now expressed in the coordinates
$\{x^a\}$ by means of the function $f$. The latter is usually simply
denoted by $g(f(x)) \equiv g(x)$. Choosing now $g=y^a$, the $a$-th
coordinate map on $V\subset N$, we find
 \be
 (f_*X)^a = X^b \frac{\partial y^a(x)}{\partial x^b}. \label{alan:push-forward_comp}
 \ee
We find that the push-forward of $X$ under the map $f$ is simply
expressed in terms of the Jacobian of $f = y(x)$. An important
property of the push-forward is that for $f:M\rightarrow N$ and
$h:N\rightarrow P$,
 \be
 (h\circ f)_* = h_* f_*\,.
 \ee

\begin{defi}
Given a map $f:M\rightarrow N$, there always exists an induced map
$f^*: T_{f(x)}^*N \rightarrow T_x^*M$, called the pull-back of $f$.
For an $X\in T_xM$, the pull-back of a 1-form $\alpha$ is given by
 \be
 f^*\alpha (X) = \alpha(f_*X).
 \ee
\end{defi}
Using (\ref{alan:formonvector}) and (\ref{alan:push-forward_comp}),
we find,
 \be
 (f^*\alpha)_b X^b = \alpha_b (f_*X)^b = \alpha_b X^c \frac{\partial y^b(x)}{\partial
 x^c}.
 \ee
Choosing now, $X = \partial/\partial x^a$, so that $X^b =
\delta^b_a$, leads to
 \be
 (f^*\alpha)_a = \alpha_b \frac{\partial y^b(x)}{\partial
 x^a}.
 \ee
Again, the pull-back of $\alpha$ under the map $f$ is simply
expressed in terms of the Jacobian of $f=y(x)$.

One can easily extend this to a definition of the pull-back of a
$p$-form $\alpha$. For $X_i\in T_xM$,
 \be
 f^*\alpha(X_1,X_2,...,X_p) = \alpha(f_*X_1, f_*X_2,...,f_*X_p)
 \ee
Now the induced map is $f^*: \Lambda_{f(x)}^pN \rightarrow
\Lambda_x^pM$ and in component form we find,
 \be
 f^*\alpha_{a_1...\,a_p}(x) = \alpha_{b_1...b_p}(y(x))\frac{\partial y^{b_1}}{\partial
 x^{a_1}} ... \frac{\partial y^{b_p}}{\partial x^{a_p}}.
 \ee
The most important properties of the pull-back of a $p$-form we will
use are,
 \be
 d(f^*\alpha) &=& f^*d\alpha, \\
 (h\circ f)^* &=& f^* h^*, \\
 f^*(\alpha\wedge\beta) &=& f^*\alpha\wedge f^*\beta,
 \ee
with $f:M\rightarrow N$, $h:N\rightarrow P$, $\alpha\in \Lambda^pN$
and $\beta\in \Lambda^qN$. These identities are not too difficult to
prove using their component form.

\renewcommand{\thesection}{\arabic{section}}

\end{document}